\documentstyle[12pt,aasms4]{article}
\doublespace
\begin{document}
\title{THE PROBLEM OF HIPPARCOS DISTANCES TO OPEN CLUSTERS:  I.  CONSTRAINTS
FROM MULTICOLOR MAIN SEQUENCE FITTING\footnotemark[1]}

\footnotetext[1]{Based on data from the ESA Hipparcos astrometry satellite.}

\author{Marc H. Pinsonneault}
\affil{Astronomy Department, Ohio State University,\\
174 West 18th Avenue,  Columbus OH 43210\\
email: pinsono@astronomy.ohio-state.edu}

\author{John Stauffer}
\affil{Harvard-Smithsonian Center for Astrophysics,\\
60 Garden St., Cambridge, MA 02138\\
email: stauffer@cfa.harvard.edu}

\author{David R. Soderblom and Jeremy R. King}
\affil{Space Telescope Science Institute\\
3700 San Martin Drive, Baltimore MD 21218\\
email: soderblom@stsci.edu,jking@stsci.edu}

\author{Robert B. Hanson}
\affil{Univ. of California Observatories/Lick Observatory,\\
University of California, Santa Cruz, CA 95064\\
email: hanson@ucolick.org}

\begin{abstract}

Parallax data from the Hipparcos mission allow the direct distance to
open clusters to be compared with the distance inferred from main sequence
(MS) fitting.  There are surprising differences between the two distance 
measurements, which indicate either the need for changes in the cluster 
compositions or reddening, underlying problems with the 
technique of main sequence fitting, or systematic errors in the 
Hipparcos parallaxes at the 1 mas level.  We examine the different 
possibilities, focusing on MS fitting in both metallicity-sensitive
\bv\ and metallicity-insensitive $V-I$ for five well-studied systems (the 
Hyades, Pleiades, $\alpha$ Per, Praesepe, and Coma Ber).

The Hipparcos distances to the Hyades and $\alpha$ Per are within 1 
$\sigma$ of the MS fitting distance in \bv\ and $V-I$, while the Hipparcos 
distances to Coma Ber and the Pleiades are in disagreement with the MS fitting 
distance at more than the 3 $\sigma$ level.  There are two Hipparcos
measurements of the distance to Praesepe; one is in good agreement with the
MS fitting distance and the other disagrees at the 2 $\sigma$ level.  The 
distance estimates from the different colors are in conflict with one another 
for Coma but in agreement for the Pleiades.  Changes in the relative cluster 
metal abundances, age related effects, helium, and reddening are shown to be 
unlikely to explain the puzzling behavior of the Pleiades.
We present evidence for spatially dependent systematic errors at the 1 mas 
level in the parallaxes of Pleiades stars.  The implications of this result 
are discussed.  
   
\end{abstract}


\section{The Problem}

Main sequence fitting is a basic tool used in the study of star clusters;
the principle behind it is also used to estimate distances to field main
sequence (MS) stars.  The Hipparcos mission (ESA 1997) has provided parallaxes 
for a number of open cluster stars, which
permits a direct determination of the distances to the open clusters which
can be compared with distances obtained from MS fitting.  There are 
surprising differences between distances obtained with these two methods;
in this paper we explore possible explanations for them.

MS fitting relies upon the Vogt-Russell theorem:  the location of a star
in the HR diagram is uniquely specified by its mass, composition, and age.
This implies that we can infer the distance of a given cluster by
comparing the apparent magnitudes of cluster stars with the absolute 
magnitudes of stars with known composition and distance.  There are 
several possible approaches.  Unevolved lower MS field stars with known 
distances or a cluster (such as the Hyades) of known distance can be used to 
construct an empirical MS.  The distance to the cluster is inferred from the 
vertical shift needed to line up the cluster MS with the empirical 
MS.  Clusters can also be compared with theoretical isochrones calibrated on 
the Sun; the latter method requires a color calibration which relates the
model effective temperatures to the observed colors.

Most nearby open clusters are close to the Sun in metal abundance, which
minimizes uncertainties in the distance scale from variations in composition.
  There is also a large database of fundamental effective temperature
measurements for stars near the solar [Fe/H], so the color calibrations  
should be relatively reliable.  The nearby open clusters also have been
extensively studied for membership, photometry, abundances, and reddening.
For all of these reasons the open cluster distance scale has not been regarded
as controversial, and evidence that MS fitting yields incorrect distances
could have significant astrophysical importance.

The Hipparcos mission has resulted in a large increase in the number of open
cluster stars with measured parallaxes.  This data allows the distance 
scale inferred
from MS fitting to be compared with the distance scale inferred from
trigonometric parallaxes.  The recently announced Hipparcos determination of 
the mean parallax of the Pleiades cluster gives the result $8.61 \pm 0.23$ 
milliarcsec \markcite{vh97a}
(van Leeuwen \& Hansen Ruiz 1997a).  This corresponds to a distance of 
$116\pm3$ pc, or a distance modulus of $5.32\pm0.06$ magnitude.  Traditional 
determinations of the Pleiades distance (e.g., \markcite{vb84} VandenBerg \& 
Bridges 1984; \markcite{s93} Soderblom
et al. 1993), comparing the cluster's main sequence to that of nearby stars,
lead to a distance modulus of about 5.6 mag ($d \sim 130$ pc; $\pi \sim 7.7$
mas).  Thus the Hipparcos parallax, being almost 1 mas larger than expected,
 suggests that the Pleiades cluster stars are systematically $\sim
0.3$ magnitude fainter than we have thought up to now.

Parallaxes for stars in other clusters have also been measured, and
the results are compared with those obtained from MS fitting in Table 1
(data taken from \markcite{phip97} Perryman et al. 1997, 
\markcite{mhip97} Mermilliod et al. 1997, \markcite{rhip97}
Robichon et al. 1997).  The standard reddening for the clusters is also
indicated, along with a notation about whether or not differential
reddening is present.  The second column lists the cluster [Fe/H] values from 
Boesgaard \& Friel (1990) and Friel \& Boesgaard (1992); we have adopted 
their abundance scale for the clusters in the present study (see Section 
4).  Mermilliod et al. 1997 and Robichon et al. 1997
concluded that there is no simple explanation for the discrepancies between
the MS fitting and Hipparcos distances, and that
all of the possible classes of solutions appeared unsatisfactory.

\begin{deluxetable}{lcccccc} 
\tablenum{1}
\tablecaption{Open Cluster Parameters} 
\tablehead{
\colhead{Cluster} & [Fe/H] & \colhead{$m-M$} & \colhead{$(m-M)_o$} & \colhead{$(m-M)_o$} & \colhead{$(m-M)_o$} & \colhead{$E(B-V)$}\\ 
\colhead{ } & & \colhead{Apparent} & \colhead{Lynga} & \colhead{Hipparcos} & \colhead{This paper} & \colhead{mag}}
\startdata
Hyades        & $+0.13$ & 3.01  & 3.01  & 3.33${\pm}0.01$ & 3.34${\pm}0.04$ & 0.00\nl
Coma Ber      & $-0.07$ & 4.49  & 4.49  & 4.73${\pm}0.04$ & 4.54${\pm}0.04$ & 0.00\nl
Pleiades      & $-0.03$  & 5.61  & 5.48  & 5.33${\pm}0.06$ & 5.60${\pm}0.04$ & 0.04\nl
IC 2602       &  & 6.02  & 5.89  & 5.84${\pm}0.07$ & \nodata & 0.04\nl
IC 2391       &  & 5.96  & 5.92  & 5.83${\pm}0.08$ & \nodata & 0.01\nl
Praesepe      & $+0.04$ & 5.99  & 5.99  & 6.24${\pm}0.12$ & 6.16${\pm}0.05$ & 0.00\nl
${\alpha}$ Per & $-0.05$ & 6.36  & 6.07  & 6.33${\pm}0.09$ & 6.23${\pm}0.06$ & 0.10\tablenotemark{a}\nl
Blanco 1      & & 6.97  & 6.90  & 7.01${\pm}0.26$ & \nodata & 0.02\nl
IC 4756       &  & 8.58  & 7.94  & 7.30${\pm}0.19$ & \nodata & 0.20\tablenotemark{a}\nl
NGC 6475      &  & 7.08  & 6.89  & 7.32${\pm}0.19$ & \nodata & 0.06\nl
NGC 6633      &  & 8.01  & 7.47  & 7.32${\pm}0.34$ & \nodata & 0.17\tablenotemark{a}\nl
Stock 2       & & 8.62  & 7.41  & 7.50${\pm}0.32$ & \nodata & 0.38\tablenotemark{a}\nl
NGC 2516      &  & 8.49  & 8.07  & 7.71${\pm}0.15$ & \nodata & 0.13\nl
NGC 3532      &  & 8.53  & 8.40  & 8.10${\pm}0.36$ & \nodata & 0.04\nl
\enddata
\tablenotetext{a}{Variable reddening}
\end{deluxetable}

We note that a second calculation of the distance to Praesepe has been
performed by \markcite{vh97b} van Leeuwen \& Hansen Ruiz (1997b), and they
find a distance modulus of 6.49$\pm$0.15 - in disagreement both with MS 
fitting and the Mermilliod et al. Hipparcos distance.  For the purposes of 
this paper we have adopted the Mermilliod distance; if we were to adopt the 
VH97b distance to the cluster we would have to add Praesepe to the list of 
clusters with a significant (2 $\sigma$) discrepancy between the MS
fitting and Hipparcos distance scales.

The first column of distance moduli in Table 1 lists the values cited as
``Lynga'' by Mermilliod et al. (1997) and Robichon et al. (1997).  We note
that these are {\it apparent} distance moduli, needing considerable (up to
1.2 mag) corrections for extinction, and cannot be directly compared with
the distance moduli $(m-M)_o$ calculated from the Hipparcos parallaxes.  
The second column in Table 1 lists the distance moduli which correspond to
the cluster distances given in Lynga's (1987) Catalogue.  These distances 
come from a variety of sources, are still scaled to a Hyades distance
modulus of 3.01 mag, and need corrections for each clusters metallicity.
  One motivation for our study is to place MS fitting distances for open 
clusters on a consistent scale.  In a paper in preparation, we have found 
that the MS fitting distances
to some of the more distant open clusters are substantially different from
the Lynga distances and in marked disagreement with the Hipparcos
parallax distances.  A second question is the precision of MS fitting
estimates; we will show that accuracy at the 0.05 mag level is possible for
well-studied systems.  Our results for the clusters studied in this paper are 
in the fourth column.  

Discrepancies between the MS fitting distances and the Hipparcos
distances could arise from several sources.  As indicated above, one
possibility is that the MS fitting distances need to be rederived on a
consistent scale.  Another possibility is that some of the basic properties 
of well-studied open clusters, such as composition, age, or reddening, need 
to be revised.  If neither of these possibilities can reconcile the distance
scales, then we are left with one of two important conclusions : either there
are fundamental problems with MS fitting or there are unrecognized systematic
errors in the Hipparcos parallaxes themselves.
  
These issues are important for other questions as well.  For example, 
recent proposed revisions to the globular cluster distance and age scales,
based on Hipparcos parallaxes of subdwarfs, rely on the same
MS fitting technique that gives rise to the puzzling distances to open
clusters (\markcite{r97} Reid 1997; \markcite{g97} Gratton et al. 1997; 
\markcite{c98} Chaboyer et al. 1998; but see also \markcite{pmtv98}
Pont et al. 1998).  

In this paper we address the essential issues raised above.  The Pleiades, 
Praesepe, and $\alpha$ Per are well-suited for a more detailed
examination.  There is good membership information and multicolor photometry
for all three; $\alpha$ Per is a system with an age comparable to that of the 
Pleiades (50 Myr vs. 100 Myr) and therefore it provides a test of age-related 
effects.  We have also examined the Coma Ber star cluster, which has a low
quoted error for its Hipparcos distance.  In a companion paper 
(Soderblom et al. 1998) we have searched for field stars with accurate 
parallaxes and anomalous positions in the HR diagram.

We begin by describing the theoretical models which we use and the open
cluster data in section 2.  In section 3 we begin with a comparison of the
Pleiades, Praesepe, and $\alpha$ Per in different colors.  We then use 
the Hyades cluster to test the zero-point of 
our distance scale, check on the shape of the isochrones in the 
observational color-magnitude diagram, and to determine the sensitivity of
distance estimates in different colors to changes in metal abundance.
We then derive distance modulus estimates at both solar [Fe/H] and the
individual abundances inferred from high-resolution spectroscopy for the 
Pleiades, Coma Ber, Praesepe, and $\alpha$ Per using several different 
methods and both \bv\ and $V-I$.  The Pleiades and Coma Ber are found to be in 
disagreement with the Hipparcos distance scale.  We discuss the sensitivity 
of our results to age, composition, and reddening in section 4, and present
evidence that the Hipparcos parallaxes may contain small-scale ($\sim$1 deg)
systematic effects $\sim$1 mas in size, large enough to cause the Pleiades
parallax discrepancy.  Our conclusions are in section 5.

\section{Method and Data}

\subsection{Theoretical Model Parameters}
Theoretical stellar models were constructed with the Yale stellar
evolution code for a range of compositions.  The nuclear reaction 
cross-sections are from \markcite{bpw95} Bahcall, Pinsonneault, \&
Wasserburg (1995).
We use the \markcite{scv95} Saumon, Chabrier, \& van Horn (1995) equation of 
state for
temperatures less than $10^6$ K; this EOS is superior to the treatment
in earlier versions of the Yale evolution codes for the conditions present
in low mass stars.  The fully ionized EOS in the Yale code was used for higher
temperatures; in numerical tests this produced a MS indistinguishable from that
obtained with the OPAL EOS (\markcite{rsi96} Rogers, Swenson, \& Iglesias 
1996).  

Model atmospheres from \markcite{k91a} Kurucz (1991a) were used as a
surface boundary condition; these are available for a range of metal 
abundances.  We also constructed models (for solar composition only) using
a grid of atmospheres kindly provided by F. Allard (see \markcite{ahas97}
Allard et al. 1997 for a review).  
\markcite{k91b} Kurucz (1991b) molecular
opacities were used for temperatures less than $10^4$ K, and OPAL opacities
\markcite{ri92} (Rogers \& Iglesias 1992) were used for higher temperatures.  
The mixing length ($\alpha=1.735$) and helium abundance ($Y=0.2704$) were 
calibrated by requiring that a solar model have the radius and luminosity of
the Sun at the age of the Sun.

We then generated models from [Fe/H]=-0.3 to [Fe/H]=+0.2 in 0.1 dex
increments.  At each metal abundance we ran a grid of masses from 0.2-1.6 
$M_{\sun}$ in 0.1 $M_{\sun}$ 
increments.  Helium abundances for non-solar [Fe/H] were obtained
by combining the solar helium abundance with an assumed $\Delta$Y
/$\Delta$Z = 2.  Isochrones were constructed at a variety of ages, to account 
for the pre-MS nature of the lower mass stars in the young clusters.  We also
constructed a set of models with solar Z but enhanced helium (Y=0.37).
The Yale color calibration \markcite{g88} (Green 1988) was used to convert 
from the theoretical (L, Teff) plane to the observational (magnitude, color) 
plane in \bv\ and $V-I$.
      
\subsection{Field Star and Cluster Data}
  For the open clusters the distance modulus estimates from
Hipparcos parallaxes are taken from the Mermilliod et al. (1997) and
Robichon et al. (1997) papers.  To perform MS fitting we restricted the sample
to stars with \bv\ in the range 0.5 - 0.9.  Distance
modulus estimates for early-type stars can be affected by the assumed cluster
age, and $V-I$ becomes a poor temperature indicator; we therefore excluded 
stars with \bv\ colors less than 0.5.  The color calibrations become 
unreliable for late-type stars; we therefore excluded stars with \bv\ colors 
greater than 0.9 from our main sequence fitting estimates.  Our choice of this
color interval also makes our distance estimates insensitive to the adopted
cluster ages.

For the photometry of open cluster stars, data was 
taken from several sources.  For the Hyades
we used cluster members (as determined by Perryman et al. 1997) which were
considered to be single stars by \markcite{ggzg88} Griffin et al. (1988).  
We restricted the
sample to stars with RI photometry (on the Johnson system) from Mendoza 
(1967).  Since the Hipparcos sample is magnitude-limited, including other stars
in common between the Perryman and Griffin paper did not result in increasing 
the sample size significantly in the color range of interest, and including 
additional stars with RI colors measured in other systems raises the 
possibility of systematic color effects.  Individual parallaxes for the Hyades 
stars were used to infer individual $M_V$ values and correct for depth effects.

For both the Pleiades and Praesepe, photometry for the upper MS stars was taken
from \markcite{m67} Mendoza (1967).  Additional data for $\alpha$ Per is from
\markcite{set85} Stauffer et al. (1985), \markcite{set89} Stauffer et al.
(1989), \markcite{pr92} Prosser (1992), \markcite{m60} Mitchell (1960),
and \markcite{pr94} Prosser (1994b).  Photometry for the lower MS stars in the 
Pleiades was taken from \markcite{st80} \markcite{st82a} \markcite{st84}
Stauffer (1980,1982a,1984),  \markcite{jm58} Johnson \& Mitchell (1958), 
and \markcite{l79} Landolt (1979).  Additional 
photometry for Praesepe was also taken from \markcite{uwd79}Upgren, Weis,
\& DeLuca, (1979), \markcite{w81} Weis (1981), \markcite{st82b} Stauffer
(1982b),  and \markcite{j52} 
Johnson (1952).  Photometry for the Coma cluster was taken from
\markcite{jk55} Johnson \& Knuckles (1955) and \markcite{m67} Mendoza (1967).

The Mendoza open cluster RI photometry is on the Johnson system, while the
other open cluster RI data is on the Kron system; both the isochrones and the
field star data are on the Cousins system.

For the reddening we adopted $E(B-V)$=0 for the Hyades \markcite{cp66}
(Crawford \& Perry 1966) and Praesepe
\markcite{cb69} (Crawford \& Barnes 1969).  For the Pleiades we adopted 
$E(B-V)$=0.04 and used individual reddenings for a small number of highly
reddened stars \markcite{cp76,b86,sh87} (Crawford \& Perry 1976, Breger 1986,
Stauffer \& Hartmann 1987).  $\alpha$ Per has patchy and variable 
differential 
reddening which is apparent in the cluster color-magnitude diagram; we
adopted $E(B-V)$=0.10 \markcite{cb74,p94} (see
Crawford \& Barnes 1974, Prosser 1994a).  We corrected the V magnitudes and
different colors for reddening as follows \markcite{a74,bb88} (Allen 1973,
Bessell \& Brett 1988):

$A_V$ = 3.12$E(B-V)$

$E(V-I)_{Cousins}$ = 1.25$E(B-V)$

$E(V-I)_{Kron}$ = 1.5$E(B-V)$

$E(V-I)_{Johnson}$ = 1.75$E(B-V)$ 

The impact of reddening on distance modulus estimates in different colors is
discussed in Section 4.  The reddening-corrected
$(V-I)_J$ and $(V-I)_K$ were converted to Cousins $(V-I)_C$ using the
transformations in \markcite{b79} Bessell (1979) and \markcite{bw87}
Bessell \& Weis (1987) respectively:

$(V-I)_C = 0.778(V-I)_J, 0<(V-I)_J<2$

(true for all Mendoza stars in this color range); and

$(V-I)_C = 0.227 + 0.9567(V-I)_K + 0.0128(V-I)_K^2 - 0.0053(V-I)_K^3$

\section {Main Sequence Fitting}

Most work on cluster distances has used \bv\ colors as an effective temperature
index.  In Figure 1 we compare the Pleiades to $\alpha$ Per and Praesepe at 
the Hipparcos distances in the observational ($M_V$ versus \bv) plane.  
Both Praesepe and $\alpha$ Per are distinctly above the Pleiades.  This
result is as disturbing as the discrepancies in Table 1 because the measured 
[Fe/H] of the three clusters are within 0.1 dex, implying that the cluster 
main sequences should be very close in this diagram (within 0.1 magnitudes).

The \bv\ color is highly metallicity sensitive, and the distances inferred from
\bv\ are therefore quite sensitive to the adopted cluster [Fe/H] values.  If
the true cluster abundances deviate from the currently accepted values then
one might expect the cluster main sequences to be closer in a 
less metallicity-sensitive index, such as $V-I$.  The clusters are compared in
$(V-I)_{Cousins}$ in Figure 2.  Praesepe is closer to the Pleiades
in this index, which suggests that part of the difference in Figure 1 is caused
by a difference in chemical composition.  However, the two cluster main 
sequences are still well separated in $V-I$ and the difference between the 
Pleiades and $\alpha$ Per is the same in both colors.  These 
figures illustrate both the magnitude of the problem and the difficulty in 
explaining it by either metallicity or age.  To quantify this problem, we
need to determine the sensitivity of distances based upon temperature 
measurements from \bv\ and $V-I$ to changes in composition.

\subsection {Theoretical Isochrones and Field MS Data}

We show theoretical isochrones for 1 Gyr populations with different
abundances
in Figure 3.  The top, middle, and bottom panels are the theoretical plane, 
$V-I$, and \bv\ respectively; the Yale color calibration was used for the
bottom two panels.  The width
of the MS is different in each; 0.5 dex in [Fe/H] 
produces a range of $\sim 0.3$ mag in the theoretical plane, $\sim 0.45$ dex 
in V/$V-I$, and $\sim 0.6$ dex in V/\bv.  Helium variations affect the 
isochrones the same in all
three planes : a 0.1 increase in Y produces a 0.28 magnitude decrease in the
locus of the main sequence.  The isochrones are nearly parallel across the
entire color range of interest.  The \markcite{aar96}
Alonso, Arribas, \& Martinez-Roger (1996) [AAR] color calibration
is an alternate method of converting from
the theoretical to the observational plane.  The AAR color calibration is 
based on application of the Infrared Flux Method, and it can be used to
derive the sensitivity of different color indices to changes in metal
abundance.  The [Fe/H] sensitivity of \bv\ in AAR is comparable to that in the 
Yale color calibration, but AAR find that $V-I$, at least in the
Johnson system, is metallicity independent : that is, that the changes in a
$V-I$ based color magnitude diagram should faithfully reflect changes in the
theoretical HR diagram.  The Hyades and Praesepe clusters
provide support for the AAR findings on the metallicity sensitivity of
$V-I$, at least for systems near solar metal abundance.

If we adopt the Yale
color calibration, a 0.1 dex increase in [Fe/H] produces a 0.12 magnitude
decrease in $M_V$ at fixed \bv\ and a 0.09 magnitude decrease in $M_V$ at 
fixed $V-I$.  For example, this would imply that the Hyades (at [Fe/H]=+0.15)
would lie 0.18 magnitudes above a solar composition isochrone in \bv\ and
0.135 magnitudes above a solar composition isochrone in $V-I$.  If the
metallicity sensitivity of AAR is adopted, then a
0.1 dex increase in [Fe/H] produces a 0.13 magnitude
decrease in $M_V$ at fixed \bv\ and a 0.06 magnitude decrease in $M_V$ at 
fixed $V-I$.  The Yale color calibration implies that a color-color
diagram in \bv\ and $V-I$ should be metallicity-insensitive because both
indices are metallicity sensitive (albeit to slightly different degrees);
the AAR results would produce a wider spread in a color-color diagram
because \bv\ is much more metallicity sensitive than $V-I$.

The local field stars span a range in [Fe/H],
with the bulk of the F and G stars in the Gliese catalog spanning a range from
-0.3 to +0.2 (\markcite{wg95} Wyse \& Gilmore 1995; see also 
\markcite{mw97} McWilliams 1997).  We therefore 
expect a significant intrinsic width to the field
star MS as well as a population of binary stars above the MS.  The
relative width of the MS in the \bv\ and $V-I$ planes can be used as a test of 
the relative sensitivities of the two color indices to [Fe/H].

The Hipparcos mission has provided parallaxes accurate to $5\%$ or
better for a significant number of lower MS stars.  There is a smaller sample 
(680 stars) with both \bv\ and $V-I$ colors; the field MS in both colors is 
compared with the models in Figure 4.  The field MS is much tighter in $V-I$ 
than in \bv, suggesting that much of the spread in Figure 4 is caused by 
atmospheric effects (the dependence of \bv\ color on metal abundance) rather 
than by interiors effects.  We therefore expect that there will be systematic
differences between the derived distances in \bv\ and $V-I$ if the adopted
[Fe/H] for the isochrone departs from that of the star, or cluster, at the
0.07 (AAR) or 0.03 (Yale) magnitude level per 0.1 dex in [Fe/H].  This
discrepancy will be in the sense that the $V-I$ distance will be longer than
\bv\ if the true metal abundance is higher and the $V-I$ distance will be 
shorter than the \bv\ distance if the true metal abundance is lower. 

\subsection{The Hyades}

The Hyades cluster provides an opportunity to check the distances derived
from MS fitting against the Hipparcos distance scale for a system with
a large number of measured parallaxes spread across a large region of the sky.
We can also compare the isochrones in the theoretical plane and in different
colors; the Hyades provides a useful check of the sensitivity of the isochrones
to changes in metal abundance because it has a metal abundance 0.1-0.2 dex
above solar.

The Perryman et al. (1997) paper provides locations for Hyades stars in the
theoretical plane as well as isochrones for both solar composition and the
Hyades [Fe/H] adopted in that study, $0.14\pm0.05$.  Our 600 Myr isochrones
for both [Fe/H]=0 and [Fe/H]=+0.14 are compared with the Perryman et al. 
isochrones in Figure 5.  For the range of $3.68 - 3.84$ in log $T_{eff}$ our 
[Fe/H]=0 isochrone is on average 0.044 magnitudes brighter than the Perryman 
et al. isochrones.  By comparison, a
zero-age MS for [Fe/H]=0 provided by Vandenberg (private communication)
is 0.032 magnitudes fainter than our 100 Myr isochrone and
for log $T_{eff}>$3.75 the Yale and Vandenberg isochrones agree to within 
0.008 magnitudes.  This comparison indicates that systematic differences
between different isochrones are at or below the 0.04 magnitude level overall
and agree to within 0.03 magnitudes near the temperature of the Sun.
 
The Hyades MS of Perryman et al. is 0.164 magnitudes brighter than
their solar [Fe/H] isochrone in this temperature interval; their isochrone
with [Fe/H]=+0.14 and solar scaled helium would be too faint to be consistent
with the data.  They were therefore
forced to a subsolar helium abundance (0.26) to reproduce the observed Hyades 
MS.  Our [Fe/H]=+0.14 isochrone with $Y=0.283$ is 0.017 magnitudes fainter 
than the Hyades MS from Perryman et al.; this implies that our models are
consistent with the Hyades having the [Fe/H] inferred from high-resolution
spectroscopy and $\Delta$Y/$\Delta$Z=2.  This result is obtained largely
because our solar composition isochrone is slightly brighter than the
Perryman et al. isochrone; we constructed isochrones with different helium
and verified that the changes in the position in the theoretical HR diagram
resulting from changes in Y and Z agree with the offsets in the Perryman
et al. paper to better than 5\%.  We obtain similar results when fitting in
the observational HR diagram using \bv\ and $V-I$.  This illustrates the
importance of small effects when inferring helium abundances based upon 
HR diagram position.  

The absolute V magnitudes of single stars and binaries in the Hyades are
shown as a function of \bv\ and $V-I$ in Figure 6.  Half of the stars
are binaries and the binaries scatter systematically above the single stars
in the color-magnitude diagram.  For comparison, the \markcite{s91} Schwan 
(1991) empirical MS for the Hyades in \bv, shifted to a distance of 3.33, is 
shown.  We also derived an empirical fit to the single star sequence in the 
$V-I$ plane which is compared with the cluster data.  The Hyades abundance 
isochrones are shifted up in the HR diagram by 0.18 magnitudes in \bv\ and 0.135
magnitudes in $V-I$ relative to a solar abundance isochrone for the metallicity
sensitivity in the Yale color calibration.  If the metallicity sensitivity
of Alonso et al. (1996) is adopted, the Hyades isochrones are 0.19 magnitudes
above the solar [Fe/H] isochrones in \bv\ and 0.09 magnitudes above the
solar [Fe/H] isochrones in $V-I$.  A slight mismatch between the shape of the 
isochrones and the empirical MS is present, and the distance modulus 
estimates from the isochrones are clearly close to the Hipparcos distance 
scale.

For each of the 35 single stars in our sample we derived a distance modulus
estimate; the average is 3.36 with a RMS deviation of 0.16.  Since the 
average error in $M_V$ is 
0.13, color errors are contributing little to the overall
scatter in the diagram.  Since the parallax errors can be correlated, the 
error in the mean scales as $N^{-0.35}$, not $N^{-0.5}$ (Lindegren 1988, 
1989); the formal error in the mean distance modulus estimate is therefore 
0.05.  Because of the nearness of the Hyades the mean distance modulus 
estimate will depend on the subset of the stars used in the comparison, and 
therefore the difference between this estimate and the cluster mean of 
3.33$\pm$ 0.01 is not problematic.

The difference ($M_V$(observed) - $M_V$(predicted)) for both colors is shown
for the single stars as a function of color in Figure 7; the difference between
the isochrones and the empirical MS is also shown.  For the Yale color
calibration the mean for \bv\ is -0.04,
implying a distance modulus of 3.32; the mean for $V-I$ is +0.05, implying a
distance modulus of 3.41.  The dispersion about the mean in both cases is
0.13, consistent with the errors in the absolute magnitudes.

The discrepancy between the distance estimates in \bv\ and $V-I$ could be 
reduced, or removed, by an increase in the adopted cluster
metal abundance; alternately, it could indicate that
\bv\ is more metallicity sensitive and $V-I$ is less metallicity 
sensitive than predicted by the models.  Since increasing the Hyades 
metal abundance would cause a disagreement with both the parallax distance to 
the cluster and the spectroscopic [Fe/H] measurements we
believe that the latter explanation is more likely.  If we adopt the 
derivatives of $M_V$ with respect to [Fe/H] from AAR the mean for the \bv\ 
distance is 3.33 (-0.03) and the mean for the $V-I$ distance
is 3.36 (0.0).  The distance we obtain by averaging the two colors is 3.34,
in excellent agreement with the Hipparcos distance.
{\it We therefore adopt the Yale color calibration at solar
metal abundance to set the zero point of the distance scale and the shape of
the isochrones and adopt the metallicity sensitivity of the Alonso et al.
(1996) color calibration:  a 0.1 dex increase in [Fe/H] produces a
decrease in $M_V$ of 0.13 magnitudes at fixed \bv\ and 0.06 magnitudes at
fixed $V-I$.}

We have added the binaries and binned the deviations between the individual
distance modulus estimates and those predicted from the isochrones.  
The resulting histograms are plotted in Figure 8.  Binaries
scatter to systematically lower distance estimates relative to single stars;
$V-I$ is more affected by this than \bv.  There are two methods that can be
used to account for binaries.  There is a clear separation between the small
mass ratio binaries and the single stars and high mass ratio binaries; the
peak in the histogram is resistant to
the presence of binaries, although there is a slight bias to lower
distance modulus estimates in $V-I$ because of its higher sensitivity to
binaries.  The peaks in the histograms of Figure 8 are at -0.05 and 0 for
\bv\ and $V-I$ respectively, which are in good agreement with the single
star averages of -0.03 and 0.  Alternately, the distance modulus estimates can
be ranked, stars more than 0.2 mag above and below the histogram peak 
excluded, and the median distance modulus estimate can be inferred.  Medians
inferred with this technique are also -0.05 and 0.0 for \bv\ and $V-I$
respectively.

We also compare the distance estimates derived from the isochrones and the
distance estimates derived from the empirical Hyades MS with the zero point
adjusted to agree with the isochrones at one solar mass for the 
different colors.  The Hyades MS needs to be corrected for the higher than
solar [Fe/H] of the cluster; the $M_V$ at fixed color for
a solar abundance MS is 0.19 magnitudes higher at fixed \bv\ and 0.09 mag 
higher at fixed $V-I$.  The empirical MS that we adopted at solar
metal abundance in \bv\ and $V-I$ respectively are therefore

$M_V = -2.75 + (4.03 + 85.7(\bv))^{0.5}$ and

$M_V = -1.976 + 13.758*(V-I) - 5.427*(V-I)^2$ (valid only from 0.55 to 0.9
in $V-I$).

For younger clusters the empirical Hyades MS needs to be corrected for age
effects (a 0.04 magnitude level effect for the Pleiades).  We took the 
difference between our 100 and 600 Myr isochrones and applied it to the
relationships given above for the Pleiades.  We stress that the distances
obtained in this manner have the same zero-point as the isochrones; we are
using the shape of the Hyades MS and not its absolute distances, and our
distance scale is therefore not tied to the distance to the Hyades (although
it is in agreement with the cluster distance as measured by Hipparcos).

\subsection {The Pleiades}
 
In Figure 9 we present histograms of the distance modulus 
estimates for the Pleiades using different techniques.  In the top panels
100 Myr solar composition isochrones were used to estimate $M_V$ from
\bv\ and $V-I$ respectively; in the bottom panels the empirical Hyades MS was
used.  The darker bins are for hotter stars and the lighter bins are for
cooler stars; a discrepancy between the mean of the two is an indication of
a deviation between the shape of the isochrone and the cluster CMD.  In the
isochrone distances cooler stars give systematically longer distance modulus
estimates at the 0.1 mag level; these color trends are removed relative to
the Hyades MS fits in the lower panels.  The good agreement between different
techniques suggests that there are small internal errors in MS fitting for
systems with good photometry.

An average of the $V-I$ distance methods yields 5.63 $\pm$ 0.02, while an 
average of the \bv\ distance methods also gives 5.63 $\pm$ 0.02.  This should
be compared with the Hyades, where a solar abundance isochrone would give
distance modulus estimates that differ by 0.1 magnitudes; this difference
is caused by a cluster [Fe/H]
0.15 dex higher than solar.  If we add the errors in quadrature this implies
that we have a 0.03 magnitude relative error in the \bv\ and $V-I$ distance
estimates, which corresponds to a 0.05 magnitude error in the photometric
[Fe/H].  \markcite{bf90} Boesgaard \& Friel 
(1990) obtained [Fe/H] = -0.034$\pm$0.024 for the Pleiades; at this metal 
abundance our \bv\ and $V-I$ distance moduli are 5.59 and 5.61 respectively.  
Our MS fitting distance to the Pleiades is therefore 5.60, and we estimate 
that errors in the metal abundance and systematic differences in the MS 
fitting technique are at the 0.04 magnitude level.

In Figure 10 we compare the Pleiades to a [Fe/H]=-0.03 100 Myr isochrone in
both \bv\ and $V-I$.  The isochrone has been shifted to a distance of 5.60.
The isochrone is an excellent fit to the cluster CMD.  We note that single 
rapid rotators are on or above the MS in $V-I$ but below it in \bv; this may
indicate that the relationship between color and temperature for these stars
is different than for slow rotators, and we therefore exclude them from
distance estimates in both this cluster and $\alpha$ Per.

A detailed binary inventory for the Pleiades has recently been published
\markcite{brn97}(Bouvier, Rigaut, \& Nadeau 1997; see also
\markcite{m92} Mermilliod et al. 1992), and we can therefore 
check for the possible impact of binary contamination on our distance 
estimates.  We compare distance
estimates for binaries, single stars, and rapid rotators in Figure 11.  As
expected, stars with very low distance estimates are binaries and for the 
high mass ratio binaries \bv\ is less sensitive than $V-I$.  The broader 
distributions for $V-I$ seen in Figure 9 are therefore a reflection of its 
greater sensitivity to binary contamination.  The techniques that we have 
applied do not appear to be affected significantly; if we use only the single 
stars a slightly higher distance of 5.65 is indicated for both colors.

To reproduce the 
Hipparcos distance of 5.33 we would require [Fe/H] = -0.25 for \bv\
and [Fe/H] =-0.45 for $V-I$.  Reproducing the Hipparcos distance to the
Pleiades by changing the metal abundance would therefore require a much
lower metallicity than that obtained by high-resolution spectroscopy;
furthermore, the distance estimates for different colors would be in strong
disagreement.  Other possibilities are discussed in Section 4.

\subsection{Praesepe}

In Figure 12 we present histograms of the distance modulus 
estimates for Praesepe using different techniques; as for the Pleiades
the top panels are relative to the isochrones (600 Myr for Praesepe) 
and the bottom panels are relative to the isochrones with the shape adjusted 
to agree with the empirical Hyades MS.  Particularly for \bv, there is a 
clear indication that the distance estimates for the uncorrected isochrones 
depend on color at the 0.1 magnitude level; this 
is more apparent for Praesepe than for the Pleiades largely because the 
Praesepe sample includes more cool stars than the Pleiades sample.  However,
the histogram peaks and medians are similar for the uncorrected isochrones
and those obtained relative to the shape of the Hyades MS. An average of the 
$V-I$ distance methods yields 6.17 $\pm$ 0.02, while an average of the \bv\
distance methods yields 6.08 $\pm$ 0.02.  This difference is significantly 
larger than our estimated relative error of 0.03.  We therefore conclude
that the difference between the two is real and indicates that Praesepe is
mildly metal-rich.

The two distance modulus estimates agree at [Fe/H]=+0.13 and a distance 
modulus of 6.25.  The \markcite{fb92} Friel \& Boesgaard (1992) [Fe/H] for 
Praesepe is +0.04$\pm$0.04, while a higher abundance is inferred by some 
other studies (see Section 4.2).  At the FB92 [Fe/H] the distance modulus is 
6.13 in \bv\ and 6.19 in $V-I$; we therefore adopt a distance modulus estimate 
of 6.16 for Praesepe (Figure 13).  This is well within the error bounds of 
the Hipparcos distance estimate of 6.24$\pm$0.12.
We conclude that Praesepe is consistent within the errors with the
Hipparcos distance measurement and that the photometry is consistent 
with it being more metal-rich (at the 0.1 dex level) than the FB92 estimate.

The dominant source of error in the distance modulus is the metal abundance of
the clusters; in general, relative metal abundances can be determined more
precisely than absolute metal abundances.  Another way of looking at the 
problem of reconciling the MS fitting and Hipparcos distance scales is 
therefore to look at relative distances in different colors and asking what 
metal abundance difference is needed to explain the results.  The Pleiades
and Praesepe are especially difficult to explain in combination.  At solar 
[Fe/H] the relative distances of these two clusters from MS fitting are 
0.45 and 0.54 magnitudes in \bv\ and $V-I$ respectively.  
  By comparison, the magnitude difference between the two clusters for the 
Hipparcos distance scale is 0.91.  If Praesepe is
metal-rich or the Pleiades is metal poor then the true difference in distance
modulus estimates will be larger than at solar [Fe/H], with \bv\ being more 
metallicity sensitive than $V-I$.  Reconciling the relative cluster distances
in \bv\ and $V-I$ by changing the metal abundances would require a metallicity 
difference of 0.35 dex and 0.6 dex respectively; both are well outside the
range of relative metal abundances reported by different investigators.

\subsection{$\alpha$ Per}

$\alpha$ Per is a young system (50 Myr), and it has a larger overall 
reddening (0.10) than the other clusters we examine and some differential
reddening.  In Figure 14 we show the distribution in distance modulus estimates
in both \bv\ and $V-I$ relative to a solar composition 50 Myr isochrone.  There
is a larger population of rapid rotators in this cluster than in the Pleiades,
and they show the same pattern (long distances in \bv\ and short distances in
$V-I$).  A subpopulation of stars at higher distances is present in both
colors, with distances systematically higher for \bv\ than for $V-I$.  This 
could be caused by variable reddening, rapid rotators with low sin i, or
contamination of the sample by non-members.  Excluding these stars only 
affects the distance estimates at the 0.02 magnitude level and does not change
the relative distances in the two colors.

There 
is a well-defined peak in \bv\ at a distance of 6.275 and the distribution for
$V-I$ is centered at the same distance; the median of the single stars is at
6.29 and 6.27 for \bv\ and $V-I$ respectively, giving average distances of
6.28 and 6.27 for the two colors.  Therefore, at solar abundance the average
cluster distance is 6.28; there is a hint of a mild metal deficiency in the
relative distances in the two colors, at the 0.02 dex level.  If we adopt the
high-resolution abundance [Fe/H]=-0.05 our average distance is 6.23; because
the error in the metal abundance is higher for this system than for the
others (0.05) the error is larger, 0.06 magnitudes.  A 50 Myr isochrone is
compared with the cluster in Figure 15.  Significantly, there is 
no evidence for a
discrepancy between the MS fitting and Hipparcos distance scales; because
both $\alpha$ Per and the Pleiades are young, this indicates that the problem
with the Pleiades is not a consequence of systematic color errors arising
from the youth of the system.

\subsection{Coma Ber}
Coma Ber is a sparse cluster with an age comparable to the Hyades; it is 
mildly metal-poor ([Fe/H]=$-0.071\pm0.020$; see \markcite{t94} Taylor 1994).  
Surveys have been undertaken to find low-mass members, and few candidates
have been found (see \markcite{rsp96} Randich, Schmitt, \& Prosser 1996 for
a discussion).  We used the Johnson \& Knuckles (1955) photometry for \bv\ and 
the Mendoza (1967) photometry for $V-I$; we note that the $V-I$ photometry for the cluster stars listed in RSP differs significantly from that in the Mendoza
1967 paper and is based on an earlier study by Mendoza.  The sample size for 
the color interval used in the other clusters is small (15 stars), so we also 
included 9 additional stars with \bv\ from 0.35 to 0.49 and $V>7.5$.  A 
histogram of the distance estimates from the isochrones is shown in Figure 
16.  There is a clear peak in the histogram for \bv\ at a distance modulus of 
4.625 at solar [Fe/H].  The Hyades MS shape yields a peak at a similar 
distance of 4.675, but with a systematic dependence of the distance estimate 
on color, i.e. the shape of the Coma Ber and Hyades MS are different for the 
hotter stars.  We therefore adopt the isochrone fit for our distance 
estimate.  Correcting for metal abundance we get a \bv\ distance modulus for
Coma Ber of 4.54; given the low quoted error in the cluster [Fe/H] we 
estimate an error of 0.04 magnitudes.  If we were to adopt a larger 
uncertainty of 0.05 dex in the cluster [Fe/H] the error in the distance
estimates for all of the clusters we have studied would rise to 0.06 
mag.  The MS fitting distance of 4.54 is a $3.4\sigma$ discrepancy with the 
Hipparcos distance if we adopt an error in the MS fitting distance of 0.04 
and $2.6\sigma$ if we adopt an error in the MS fitting distance of 0.06.

The behavior of the cluster in $V-I$, however, is puzzling.  There is no
well-defined peak for the cool stars, and the hotter stars concentrate at a
distance (5.1) well above either the \bv\ or the Hipparcos distance.  This
can be traced to the temperature scale for the two colors; the $V-I$ colors
for the cluster F stars imply temperatures significantly hotter than the \bv\
colors.  The isochrones are brighter at the hotter temperatures, causing
higher distance modulus estimates for these stars (smaller $M_V$ implies
larger m-M).  We compare the temperatures from the isochrones in the two
colors to those obtained by \markcite{bo87} Boesgaard (1987) in a study of
lithium in F stars in Figure 17.  Boesgaard estimated temperatures from \bv,
Stromgren photometry, and also measured spectroscopic temperatures; her
temperature scale is in excellent agreement with the \bv\ temperature scale in
the isochrones and in significant disagreement (at the 200-300 K level) with
the $V-I$ temperature scale for the F stars.  Adopting the hotter temperature 
scale implied by the $V-I$ colors would raise a series of problems : the 
lithium dip in Coma would be at hotter $T_{eff}$ than for other clusters, it 
disagrees with spectroscopic temperature estimates and those from Stromgren 
photometry, and large internal
variations in the derived iron abundances for cluster stars would result.
We have no explanation for this problem, and reobserving the cluster stars in
$V-I$ and IR colors would be useful to understand the problem.  We therefore
do not use the $V-I$ distance to the system, and our MS fitting distance to
this cluster must be taken with this discrepancy in mind.  The cluster is
compared to isochrones in both of our colors in Figure 18.

\section{Discussion}

The results of Section 3 indicate that it is the Hipparcos distance to the 
Pleiades which is in the most serious conflict with MS fitting.  In all of the
other systems except Coma Ber, MS fitting in different colors yields distance 
results that are consistent with one another, normal helium, and [Fe/H] 
values from high resolution spectroscopy.  Coma Ber may have an equally serious
disagreement, but the unusual behavior of the cluster in $V-I$ suggests that 
other problems may be contributing to the discrepancy for it.  We therefore 
examine in turn the various possible mechanisms that could reconcile the 
cluster distance scales for the Pleiades; in all cases we believe that they 
cannot do so.  In a companion paper we show that the same conclusions result 
from an examination of nearby field stars \markcite{s98} (Soderblom et al. 
1998).  We then proceed to an analysis of the Hipparcos parallaxes for the 
Pleiades, and show that there are indications of possible systematic errors 
that could be the origin of the discrepancy.

The calculations that we have presented are standard stellar models.
We have therefore not included physical processes such as gravitational 
settling, rotational mixing, magnetic fields, internal gravity waves, or 
mass loss, which are surely present.  There are strong reasons for
believing that these nonstandard effects will not influence the distance
scale, although they could be potentially important for other issues.  
The single most important reason is the youth of the clusters that we have
examined; detailed nonstandard calculations predict little, if any, effect
for ages as young as the Pleiades.  In addition, any such process would have
to affect stars with a wide range in masses to a similar extent and be 
different among different clusters to explain the pattern that we see.

Gravitational settling is minimal in young systems such as the Pleiades,
and the degree to which helium and heavy elements sink depends strongly on
the convection zone depth and thus the stellar mass.  For example, helium
and heavy element diffusion are a 10\% fractional effect in the Sun, which is
almost 50 times older than the Pleiades.  The observed cluster lithium
abundances require a mild envelope mixing process, and models with rotational
mixing that are consistent with the lithium data predict little or no deep 
mixing (Pinsonneault 1997).  In addition, the observed range in rotation rates
in clusters is large, and any extra mixing would produce a spread in MS
properties rather than a uniform shift in the distance estimates.  Other
physical processes could affect the results, but they are still subject to
a variety of observational constraints which make a large effect unlikely.

We have compared different standard model calculations, and the zero-point
offset is small (0.01-0.03 mag for stars between 5600 and 7000 K, for
example).  The systematic errors in the standard model distance estimates is
therefore also too small to explain the results that we have obtained.  We
now discuss age, composition, and reddening effects.

\subsection{Age and Stellar Activity}
It is well-known that many young stars are heavily spotted; this could 
influence the color-temperature relationship and therefore the distance 
estimates for young systems such as the Pleiades and $\alpha$ Per.  In 
Figures 1 and 2 we compared these two clusters at the Hipparcos distances in 
our two colors; the Pleiades is clearly anomalous with respect to $\alpha$ 
Per if the Hipparcos distance scale is adopted.  Since $\alpha$ Per is 
younger and has a
larger population of rapid rotators, if anything $\alpha$ Per should be more
anomalous than the Pleiades if our MS fitting age estimates were in error
because of activity.  We note that similar conclusions can be obtained by
comparing young and old field stars (Soderblom et al. 1998).  The 
narrow width of the Pleiades MS also indicates that a wide range in stellar 
activity does not produce a significant effect on the color-temperature 
relationship.  For all of these reasons we reject the idea that youth
is responsible for the difference between the distance estimates.

Another possibility is that activity could be influencing the Pleiades [Fe/H],
which has been derived from LTE model atmospheres.
If such a phenomenon were at play, it might lead to derived abundances being 
a function of line strength due to the direct effect of activity on the 
stronger lines formed at smaller depths in the photosphere. We have a 
number of high resolution spectra of Pleiades members that was originally
obtained to study lithium abundances.  We have analyzed the \ion{Fe}{1} 
data in the cool Pleiades dwarfs and find no such [Fe/H]-line strength 
correlation.  This does not exclude such a real correlation, though, given the 
influence of damping, which is adjusted to enforce such a lack of 
correlation.  To the extent that our damping assumptions seem quite reasonable 
compared to numerous other fine spectroscopic analyses, and are consistently 
applied in both the stellar and solar analyses to 
yield line-by-line [Fe/H] values, the analysis suggests any such trends are 
not substantial.  Regardless, any systematic error in the 
inferred mean [Fe/H] is greatly mitigated by the fact that the damping 
adjustments enforce 
consistency with the weaker lines, which are formed at deeper depths, and thus 
presumably are more immune from the direct effects of chromospheric activity.
Activity in very young stars can manifest itself in the form of an 
effective veiling continuum.  Such behavior would presumably weaken the 
line absorption, thus leading to {\it underestimated} line strengths and,
hence, abundances.  Detailed NLTE line formation calculations to determine 
how the active Pleiades dwarfs' Fe and other metal abundances might be 
affected by activity, spots, convective flows, {\it etc.\/} would be of 
interest, but are unlikely to produce large errors for the reasons discussed 
above.

\subsection{Heavy Metals}

\subsubsection{The Cluster [Fe/H] Scale}
Homogeneous Fe abundances are available for the Pleiades, Praesepe, and 
$\alpha$ Per from the work of Boesgaard and collaborators.  Independent 
modern fine analyses of these clusters (and a few others) by other 
investigators are available 
for comparison with their work.  All the studies considered here derive 
self-consistent solar Fe abundances with which the stellar values are 
normalized.  Such a careful differential procedure can greatly reduce errors 
introduced by varying assumptions concerning the solar Fe abundance, model 
atmospheres, $gf$ values, {\it etc}. 

\markcite{bbr88} Boesgaard {\it et al.\/}~(1988) determine a mean Pleiades 
iron abundance of [Fe/H]$=-0.03$ from analysis of 17 F stars.  The mean 
star-by-star reddening they use is essentially identical to the value we have 
adopted.  \markcite{b88}Boesgaard (1989) determined a ``best'' Pleiades 
abundance by analyzing new data for 8 Pleiads; the result was 
[Fe/H]$=+0.02$.  Boesgaard \& Friel (1990) used new data for 12 of the same 
stars in Boesgaard {\it et al.\/} to find a mean [Fe/H]$=-0.03$.  The single 
datum standard deviation in all these studies is ${\sim}0.07$ dex.  The 
1${\sigma}$ level error in the mean is 0.02-0.03 dex, so the internal 
statistical uncertainties appear to be small.  
\markcite{ccc88}Cayrel {\it et al.\/}~(1988) derive a mean Pleiades [Fe/H] of 
$+0.13$ from analysis of four Pleiades dwarfs, three of which are 
significantly cooler (mid G) than the Boesgaard F stars.  The standard 
deviation is 0.10 dex, which is somewhat smaller than their estimated 
individual errors; the error in the mean is ${\sim}0.06$ dex.  The 
${\sim}0.1$ dex offset between the Cayrel and Boesgaard values 
is representative of uncertainties in reddening (which enters via photometric 
$T_{\rm eff}$ determinations by Boesgaard), the $T_{\rm eff}$ 
determinations (the Cayrel values are based on H$\alpha$ profiles), and other 
details.  The Cayrel result is 
consistent with \markcite{e86}Eggen's (1986) inference from narrow band 
photometry that the Pleiades [Fe/H] is near the Hyades value  

In order to increase the sample of Pleiades stars with [Fe/H] determinations, 
some of us (\markcite{k97}King {\it et al.\/}~1997) have used high quality 
Keck spectra of two slowly rotating very cool ($T_{\rm eff}{\sim}4500$ K) 
Pleiades dwarfs to derive Fe abundances.  Our $T_{\rm eff}$ values are 
spectroscopic determinations from balancing the abundances as a function of 
excitation potential, and the normalized abundances are derived by comparison 
with similarly analyzed solar data on a line-by-line basis.  The mean 
abundance is [Fe/H]$=+0.06$, with estimated errors in the mean of perhaps 
0.05 dex.  While comparison of the different studies indicates there may be 
systematic errors at the 0.1 dex level, we regard this (dis)agreement to be 
quite satisfactory given the ${\sim}2000$ K range in $T_{\rm eff}$, the 
disparate sources of data, and distinct methods used to derive $T_{\rm eff}$.  
While a slightly sub-solar Fe abundance is often assumed 
for the Pleiades based on the Boesgaard \& Friel results, the totality of the 
high-resolution spectroscopic evidence may be more consistent with a slightly 
super-solar value; our photometric [Fe/H] is consistent with solar 
[Fe/H].  Therefore, if anything the data suggest a distance modulus estimate 
larger than our MS fitting value rather than smaller.   

Fe abundances for Praesepe F dwarfs have been derived by \markcite{bb88}
Boesgaard \& Budge (1988), Boesgaard (1989), and Friel \& Boesgaard (1992). 
The resulting values are $=+0.14$, $+0.10$, and $+0.05$, with star-to-star 
scatter of 0.06-0.07 dex, and mean uncertainties of 0.03-0.04 dex; again, the 
internal precision is good.  The zero-reddening assumed in their 
$T_{\rm eff}$ determinations is identical to our assumption.  Other detailed 
studies of numerous Praesepe stars comparison are lacking.  Analysis of the 
primary component of the Praesepe SB2 KW367, a mid-G star which is 
significantly cooler than the Boesgaard F stars, by \markcite{kh96}King \& 
Hiltgen (1996) yielded [Fe/H]$=+0.01$ with an uncertainty near 0.05 dex.  
Again, systematic errors at the 0.1 dex are indicated by this limited 
comparison.  Combined with the above results, we see that [Fe/H] for Praesepe 
is 0.00-0.15 dex larger than for the Pleiades, with a preference for the 
lower middle of this range.  The results inferred from MS fitting are 
consistent with the upper end of the range.  

Boesgaard {\it et al.\/}~(1988), Boesgaard (1989), and Boesgaard \& Friel 
(1990) derived Fe abundances in $\alpha$ Per F stars.  The mean [Fe/H] values 
are $-0.02$, $+0.00$, and $-0.05$.  The $\alpha$ Per Fe abundance seems 
nearly identical to the 
Boesgaard Pleiades estimate.  The star-to-star scatter in the larger $\alpha$ 
Per samples is 0.08-0.09 dex; mean uncertainties are ${\sim}0.04$ dex.  The 
mean of the individual $\alpha$ Per reddening values employed by Boesgaard is 
${\sim}0.03$ dex lower than the single value adopted here.  This difference 
might require a 0.05-0.10 dex increase in [Fe/H] for consistency with our 
assumptions. \markcite{bls88}Balachandran {\it et al.\/}~(1988) determined 
Fe abundances in a very wide range (F to K type) of $\alpha$ Per stars.  The 
mean abundance of the stars not considered by them to be non-members is 
[Fe/H]$=+0.04$ with a star-to-star scatter of 0.14 dex; the mean internal 
error is only 0.02 dex. Their assumed reddening is identical to our value.  
The results of Boesgaard {\it et al.\/}~and Balachandran {\it et al.\/}~agree 
to within 0.1 dex, but when adjustment is made for the slightly different 
reddening assumptions, the agreement is within a few hundredths of a dex if 
not exact.  Our photometric [Fe/H] is slightly sub-solar, at the 0.01-0.02
dex level.  It thus appears that the Fe abundance of $\alpha$ Per is not 
significantly larger than for the Pleiades. 

In sum, internal errors in the Fe abundances of main sequence Pleiades, 
Praesepe, and $\alpha$ Per stars derived from careful homogeneous analyses 
employing high quality data lead to uncertainties of only 0.05-0.10 dex in 
relative cluster 
abundances.  We have seen that systematic effects due to errors in reddening, 
differences in the analysis methodology, {\it etc.\/} may approach 0.15 dex.  
These are small compared to the offset needed to explain the Hipparcos-based 
M$_V$ values for the Pleiades.  Barring fundamental failure or incompleteness 
in our understanding of spectral line formation and stellar atmospheres, the 
extant data suggests that the Fe abundances of the Pleiades, Praesepe, and 
$\alpha$ Per are within $\sim0.10$ dex of each other.  We might caution, 
however, that the abundances of other important atmospheric opacity 
contributors ({\it e.g., Mg and Si}) are, unfortunately, unknown.  

\subsubsection{Photometric Constraints and the Binary Distance to the 
Pleiades}
There are other factors that make a large error in the Pleiades [Fe/H] 
unlikely.  Colors that incorporate an infrared band are less sensitive to
metallicity than \bv.  The figures in the previous section indicate clearly
that the shift in the cluster distance modulus is the same for different
color indices; the Pleiades must be intrinsically subluminous if the
revised distance estimate is correct.  The deviations from the high-resolution
[Fe/H] values for the Pleiades are both large and inconsistent from color to 
color.  The spectroscopic binary HD 23642 also provides a distance of
$5.61\pm0.26$ consistent with MS fitting, albeit with a large error 
\markcite{gia95}(Giannuzzi 1995).

\subsection{CNO Abundances}

Carbon, nitrogen, and oxygen can affect stellar structure in ways
other elements do not; are they anomalous in the Pleiades?  As part
of his thesis, King (1993) examined the oxygen abundances of stars
in several clusters over a broad range of age.  The [O/H] for the Pleiades
was found to be higher than for Praesepe (+0.29 and +0.02 respectively, with 
errors in the mean of 0.08 for both).  However, the trustworthiness of
abundances (such as these) derived from the high excitation 7774 \AA 
\ion{O}{1} lines is a matter of some debate.  Besides possible large data and
analysis differences between various studies (e.g. King \& Hiltgen 1996),
there may be significant abundance corrections due to non-LTE effects
on line formation in stellar atmospheres (see Garcia Lopez et al. 1995).  
Unfortunately, systematic errors of 0.3 dex in the cluster O abundances
derived from high-excitation lines remains plausible.  In any case, the King 
results would act to make 
the Pleiades more metal-rich and therefore require a higher distance modulus 
estimate.  Detailed abundance studies would be useful, but deviations from the
solar mixture would need to be very large to have a significant impact on the
luminosity of the MS.

\subsection{Helium}

The initial solar helium abundance can be inferred from theoretical solar 
models by the requirement that the model have the solar luminosity at the age 
of the Sun.  Modern evolution codes give estimates for the initial solar $Y$ in
the range $0.26 - 0.28$; the best solar models of Bahcall, Pinsonneault, \&
Wasserburg (1995) had $Y = 0.272$ and $Y = 0.278$ with and without 
gravitational settling respectively.  A comparison of theoretical stellar 
models with the Hipparcos main sequence of the Hyades by Perryman et 
al. (1997) yields $Y = 0.26\pm0.02$; for comparison, the solar $Y$ in that 
study was 0.266 and the solar-scaled helium for the cluster would be 0.28.
This agreement between the Sun and Hyades was anticipated and
reinforces the notion that stars formed in the current epoch have
similar helium abundances.

Nevertheless, we consider what range of $Y$ would be needed to drop
the Pleiades main sequence by 0.3 mag, and that value is about
$Y = 0.37$.  Such a high value of $Y$ for the Pleiades would imply a drastic
revision of chemical evolution models and, by extension, would raise the
possibility that other clusters might have similar anomalies.  MS fitting
would therefore require knowledge of both the metal and helium abundances;
since helium can only be directly observed in young systems this would make 
MS fitting unreliable at the 0.3 magnitude level for the majority of clusters.

We believe that this question is best answered by direct measurements of the
helium abundance in HII regions and massive stars.  We begin with a discussion
of the literature on helium abundances; we have also obtained data on the 
relative helium abundances in the Pleiades and $\alpha$ Per.  Neither the field
star data nor our Pleiades spectra are consistent with significant variations 
in the initial helium abundance from the solar value.

Ignoring a deviant few percent of field stars, \markcite{n74}Nissen's (1974) 
study revealed no intrinsic scatter in $Y$ greater than ${\sim}10$\% 
(compared to the 30-40\% deviation required by the Pleiades stars) in nearby 
main-sequence field B stars.  \markcite{gl92} Gies \& Lambert (1992) found
helium abundances consistent with both the Sun and the Orion nebula for a 
sample of 35 B dwarfs; 4 B supergiants in that sample were found to have
anomalously high helium abundances.  There is evidence that evolutionary 
effects are responsible for helium enrichment in the most massive stars
(see \markcite{mc94}Maeder \& Conti 1994, \markcite{lu96}Lyubimkov 1996, 
\markcite{pin97}Pinsonneault 
1997 for reviews), so helium abundances from MS O stars and massive 
supergiants may not be reliable indicators of the initial $Y$.  The B star
field data and the Orion nebula abundances are therefore our best test for
the range in helium abundance at solar metal abundance, and they are 
consistent with only small variations in the initial helium abundance.

For Galactic clusters, however, the picture is less clear. 
\markcite{ss69}Shipman \& Strom (1969), \markcite{ps73}
Peterson \& Shipman (1973), \markcite{n76}Nissen (1976), and  
\markcite{l77}Lyubimkov (1977) found evidence for 20\%-30\% variations in
$Y$ among young associations, including some systems with significantly
lower $Y$.  Lyubimkov suggested an increasing He abundance with 
{\it increasing\/} age amongst the young clusters/associations studied,
a conclusion not supported by the subsequent field star work of Gies \& 
Lambert.  

\markcite{p79}Patton (1979) determined He abundances of 60 stars in 8 young 
clusters and associations.  She noted that her initial abundances displayed 
a range in $Y$ of about 25\%, and that this could not be explained by the 
the usual error sources; she also called attention to a correspondence between 
He abundance and cluster age.  However, Patton shows that binarity may be 
responsible for observed cluster-to-cluster He abundance dispersions, and 
the notably low He abundances (observed by others too) seen for a few stars 
within a given cluster/association.  Eliminating {\it suspected\/} (but not 
positively identified) binary systems from her analysis results in cluster 
He abundances which are identical to within the uncertainties.  This 
highlights the need for secure knowledge of very fundamental stellar 
parameters (e.g., binarity) before reliable He abundances can be derived. 

With this muddled picture of main sequence stellar He abundances, one may 
wonder if the Pleiades He abundance could be abnormal.  Both the Pleiades 
and $\alpha$ Per are young enough to have B stars, and their helium can be
directly measured.  The $Y$ values from Lyubimkov (1977) agree to within 
${\Delta}Y{\sim}0.015$, which is well within the uncertainties; the Pleiades 
and $\alpha$ Per $Y$ value is 0.04 larger than the corresponding field star 
value, but the uncertainties are comparable to this offset.  \markcite{kp86} 
Klochkova \& Panchuk (1986) also derived B-star He abundances in both the 
Pleiades and $\alpha$ Per.  They claim to find no difference between the mean 
abundances that is larger than the uncertainties.  However, this conclusion 
is not clear to us from the abundances listed in their Table II, which 
do demonstrate quite a very large difference.  Unfortunately, only two 
Pleiades stars are included in the analysis.  Therefore small number 
statistics and the possible effects of binarity make assessment of this 
difference quite difficult.  We attempted a final comparison using the 
``field'' stars from Nissen (1974).  This sample includes four $\alpha$ Per 
stars, and two stars (HR 5191 and 7121) which are 
suggested members of the purported Pleiades supercluster.  The mean $Y$ value 
is only 0.03 larger for the Pleiades field stars than for the $\alpha$ Per 
stars; the uncertainties are probably not much smaller than this difference. 

To investigate the possibility of a non-standard helium abundance in the
Pleiades \markcite{fk98} Fischer \& King (1998) observed MS B stars in
$\alpha$ Per and Pleiades to differentially compare the helium abundances.
Preliminary analysis of the lines strengths for six He lines 
suggests that the cluster He abundances are
identical within an uncertainty of 15\%.  Any real difference appears to be
in the opposite sense of what is needed to make the Pleiades underluminous:
the Pleiades line strengths are, if anything, consistently smaller than the
$\alpha$ Per counterparts.

\subsection {Reddening and Systematic Errors in the Photometry}

Reddening will tend to make a cluster MS fainter at a given color.  If the
reddening is increased the inferred distance modulus will therefore increase.
The effect can be roughly estimated as follows : in the color interval that
we are using for MS fitting the derivative of $M_V$ with respect to both 
\bv\ and $V-I$ is $\sim$5.  The extinction $A_V$=3.12$E(B-V)$ and 
$E(V-I)_K$=1.5$E(B-V)$.  Adding these effects together an increase in
$E(B-V)$ of 0.10 magnitudes would increase V at fixed \bv\ and fixed $V-I$
by 0.188 (0.5 mag from a shift of 0.1 in \bv\ - 0.312 mag from extinction)
and 0.438 (0.75 mag from a shift of 0.15 in $V-I$ - 0.312 mag from
extinction) magnitudes respectively.  The relative distances inferred
by the two colors can therefore be affected if the reddening is incorrect.
In addition the [Fe/H] abundances derived for cluster stars are sensitive
to T$_{eff}$, and an increased reddening would imply a higher [Fe/H] for a
given equivalent width (therefore further increasing the distance modulus).
Other colors, such as $R-I$, will be less reddening-sensitive.

Neither the Hyades nor Praesepe show any evidence for reddening along the line
of sight; increasing the reddening estimate for the Pleiades would
worsen the discrepancy with the Hipparcos distance modulus estimate.
Even changing $E(B-V)$ from 0.04 to 0 would only decrease the distance modulus
by 0.08 magnitudes.  The reddening estimates for the Pleiades have been
derived for a wide range of masses and from different techniques; Crawford
and Barnes used Stromgren photometry to estimate $A_V$ for B, A, and early
F stars in the Pleiades and Praesepe, Prosser and Stauffer used M dwarfs in
the same clusters, and Breger used polarization measurements in the Pleiades.
We conclude that reddening is not a significant source of uncertainty
in distance estimates for the Pleiades.  Multicolor distance measurements of
the type performed in this paper could be a useful check on the reddening for
more heavily obscured systems.

Another possibility is that systematic errors in the photometry could cause
errors in the distance estimates.  For the color range that we are considering,
the slope of the MS is $\sim$5; this would require a systematic error
of 0.06 magnitudes in \bv\ to reconcile the Pleiades distance scales, which is
unreasonably large.  The size of the systematic errors can be 
constrained by comparing spectroscopic temperature estimates with those 
based upon colors.  In the case of Coma Ber, for example, it appears that 
spectroscopic temperature estimates are in agreement with the \bv\ colors 
of F stars but not with the $V-I$ colors.  We note that the slope of the MS 
in $V-I$ is steeper for F stars than for the cooler stars, and that 
systematic errors in the $V-I$ photometry might explain the puzzling behavior
of Coma Ber.  We have attempted whenever possible to rely upon a single
source for photometry in a given color for a given cluster.  Even in
the case of the $V-I$ data we see no evidence of systematic differences
between the location on the color magnitude diagram of stars with colors 
converted to the Cousins system from the Kron system and those converted 
to the Cousins system from the Johnson system.

For the Pleiades, independent studies (Section 2.2) give consistent 
photometry for individual stars at the level of the quoted errors 
(0.01 - 0.02 mag).  The 0.3 mag discrepancy between the Hipparcos and  MS 
fitting distance distance modulii is much too large to be explained by 
systematic errors in the photometry.  High-resolution spectroscopy of the
Pleiades is consistent with the observed colors, and the reddening is 
small.  For systems with higher reddening, 
however, care must be taken when converting between different photometric 
systems; the Johnson, Cousins, and Kron system I bands have different 
effective central wavelengths and therefore different reddening corrections.

\subsection{Systematic Errors in the Hipparcos Parallaxes}

The final possibility is that the Hipparcos Pleiades parallaxes may contain
previously undetected systematic errors.  If the MS fitting result 
$m-M = 5.60$ does indeed give the correct Pleiades
distance, then a systematic zero-point error would need to approach the 1 mas 
level to produce the discordance with the Hipparcos results.
Such an error seems impossibly large, in view of the extensive tests
\markcite{ar95,ar97}
(Arenou et al. 1995, 1997) demonstrating the global zero-point error of the
Hipparcos parallaxes to be smaller than 0.1 mas.  However, global tests
have little power to reveal effects occurring on the small angular scale 
($\sim 1 \deg$) of the Hipparcos spatial correlations (see below).
Indeed, the Hipparcos parallaxes of stars in open clusters such as the 
Pleiades represent the first real opportunity to test for systematic effects
on small angular scales.  One might well argue that it would only be prudent
to consider the Hipparcos cluster results as the first direct tests for
small-scale zero-point errors, rather than as cluster distance measurements.

The Hipparcos Pleiades parallax (van Leeuwen \& Hansen Ruiz 1997a) is based 
on measurements of 54
cluster members, ranging
in $V$ from 2.8 to 11.5 within $5\deg$ of the cluster center, so it 
represents a fairly broad sampling of the cluster.
Because Hipparcos observed
widely separated ($\sim 58\deg$ apart) star fields simultaneously, the 
parallaxes are inherently on an absolute scale over the whole sky.  Over small
regions of the sky ($\lesssim 2\deg$), however, the astrometric results are 
positively correlated because neighboring stars (within the $0.9\deg \times \
0.9\deg$ Hipparcos field of view) tended to be observed on the same great 
circles the satellite swept out over the sky (Lindegren 1988, 1989).  A 
comprehensive discussion 
of the Hipparcos mission and data reductions is given in Volumes 1--3 of the 
Hipparcos Catalogue \markcite{ESA97}(ESA 1997).  The spatial correlations 
may significantly impact the astrometric results for star clusters, whose 
angular size is of the same order as the Hipparcos correlation scale.  To 
account for this, \markcite{vh97a} van Leeuwen \& Hansen Ruiz (1997a) 
re-calculated the Pleiades mean parallax from the intermediate Hipparcos data.
For this paper, one of us (R.B.H.) has re-examined the individual Pleiades 
parallaxes from the Hipparcos Catalogue.

Moreover, besides the spatial correlations, there is a different type of 
correlation affecting the Hipparcos results -- the statistical correlations
among the five astrometric parameters (position, proper motion, and parallax),
arising from the imperfect distribution of Hipparcos observations on the sky
over time.

In classical parallax work (cf. \markcite{va75} Vasilevskis 1975), the time 
distribution of 
observations over a star's parallactic ellipse is controlled to maximize the 
parallax factors and minimize the correlations between position, proper motion,
and parallax.  This is easy to achieve from the ground, but Hipparcos could 
not do this because of the limited span of observations and the pattern of 
scans of the sky, as explained in Section 3.2.4 (pp. 321-325) of the Hipparcos 
Introduction (ESA 1997, Vol. 1).  Figures 3.2.42 to 3.2.61 of that work 
illustrate the patterns of the correlations over the sky; Figure 3.2.66 
(p. 363) shows histograms of the 10 correlations.  The RMS values are $\sim 
0.2$, and large areas of the sky show correlations averaging 0.4 or more in 
size.  It must be emphasized that these correlations are substantially larger
than would be considered acceptable in ground-based parallax observations.

For parallax work, the most important correlation is $\rho_\alpha^\pi\ $,
between parallax and right ascension (Field H20 in the Hipparcos Catalogue).
This is because, over most of the sky, most of the extent of the parallactic
ellipse is in right ascension.  The Hipparcos $\rho_\alpha^\pi$ correlation 
is shown in Fig. 3.2.44 of the Hipparcos Introduction.  Large values of
$\rho_\alpha^\pi$ were caused in certain areas of the sky by the unfortunate 
circumstance of unequal observations on both sides of the Sun, as discussed on
p. 325 of the Hipparcos Introduction.  

This happens to impact the Pleiades particularly badly.  The mean value of
$\rho_\alpha^\pi$ near the Pleiades center is +0.4; this is at the 96th 
percentile in the histogram in Fig 3.2.66. The question this raises is whether
this large correlation, caused by the time distribution of Hipparcos 
observations of the Pleiades stars, has any effect on the parallax values.  

We stress again that this is a different effect from the spatial correlation
that exists because Hipparcos astrometric data over small ($\sim 1 \deg$) 
areas of the sky are not fully independent measurements.

In Figure 19 we plot parallax vs. the correlation $\rho_\alpha^\pi$ for 49 
Pleiades members verified by proper motion, radial velocity, and position in 
the color-magnitude diagram. (Mermilliod et al's 51 stars and van Leeuwen
et al's 54 are virtually the same set as these; we rejected several additional
stars on account of problems noted in Fields H30 and H59 of the Hipparcos
Catalogue.)  This plot shows several interesting things.

The filled symbols are 12 bright ($V < 7$) stars within $\sim 1 \deg$
of the cluster center with correlations $\rho_\alpha^\pi \geq +0.34$ 
(the mean value for the whole sample).  Due to the spatial correlation effect,
these 12 stars all have nearly the same parallax (mean 8.86 mas, RMS 
dispersion 0.45 mas; $\chi^2$ too small at the 0.995 significance level).  
Because Hipparcos' errors are smallest for bright stars, these stars carry 
much of the weight of the Pleiades parallax.

There is a clear trend (slope) of parallax vs. $\rho_\alpha^\pi$ 
correlation; a weighted least-squares solution gives a slope of 
$+3.04 \pm 1.36$ mas per unit correlation.  The solid line in Fig. 19 is this 
slope, run through the mean point (+0.34,+8.53).  The dashed lines show 
$\pm1\sigma$ slopes.  The intercept at zero correlation is 
$\pi = 7.49 \pm 0.50$ mas, quite consistent with the MS fitting distance.

Figure 20 plots parallax vs. distance from the cluster center.  The filled 
symbols are the same 12 bright stars with high $\rho_\alpha^\pi$ as in Fig. 
16.  The open symbols are the 15 stars with $\rho_\alpha^\pi < +0.25$,
with no restriction on magnitude or distance.  The two sets of stars barely
overlap because the brightest stars in the Pleiades are highly concentrated
to the cluster center.  The low-correlation stars lie farther from the Pleiades
center and show a much larger parallax scatter, reflecting (a) the larger 
errors for fainter stars and (b) the lack of spatial correlations on scales 
$\gtrsim 1 \deg$.  
Moreover, their mean parallax is smaller (reflecting the slope
discussed above).  For the 15 stars with $\rho_\alpha^\pi < +0.25$, the 
weighted mean parallax is $7.46 \pm 0.43$ mas.  The RMS dispersion is 1.66 mas,
consistent with the published parallax errors.

This exercise is not intended to be a definitive re-determination of the
Pleiades parallax; that would require going back to the intermediate
Hipparcos data as per van Leeuwen et al (1997), and exploring the effects
of both the $\rho_\alpha^\pi$ and the spatial correlations at that level.
However, it is quite clear that (a) small-angular-scale systematic effects at
the 1 mas level are present in the Hipparcos Pleiades parallaxes; (b) these
effects are related to the high values of the $\rho_\alpha^\pi$ correlation
near the cluster center; (c) the bright stars within $\sim 1\deg$ of the 
center, which carry most of the weight of the mean parallax, are the most
severely affected; and (d) the stars with lower $\rho_\alpha^\pi$ correlations,
far enough ($\gtrsim 1 \deg$) from the center to be unaffected by the spatial 
correlation, have smaller parallaxes, consistent with the MS fitting distance.

We also looked for effects of the $\rho_\alpha^\pi$ correlation in the Hyades,
Praesepe, $\alpha$~Per, and Coma~Ber clusters.  In Figures 21--24 we present 
the parallax vs. correlation plots for those clusters.  The Hyades, Praesepe, 
and $\alpha$~Per clusters also have large values of $\rho_\alpha^\pi$, but the
the slope (d$\pi$/d$\rho$) present in the Pleiades data does not occur in 
these clusters, where the MS fitting distances and the Hipparcos distances 
are in good agreement.  The data for Coma~Ber do show a slope 
d$\pi$/d$\rho \ = \ -4.0 \pm 2.1$ mas, but the range of $\rho_\alpha^\pi$ is
small, and the mean is near zero.

\section{Resolution of the Problem}

The Hipparcos distances to the open clusters can be regarded as either
a test of the theory of stellar structure and evolution or as a test of the
parallaxes themselves.  To distinguish between the two it is necessary to
determine what the errors in stellar interiors-based cluster distances are.
We have performed a detailed multicolor analysis of 
the distances to the nearby open clusters, and verify that MS fitting can be 
performed to a precision of order 0.05 magnitudes.  With the exception of 
Coma Ber, distance
estimates from \bv\ and $V-I$ colors can be used to get photometric [Fe/H]
values accurate at the 0.05 dex level, and these estimates are in good
agreement with those obtained from high-resolution spectroscopy.  There is
a small zero-point shift, of order 0.04 dex, between our photometric
abundance scale and that of Friel and Boesgaard; if we adopted our zero-point
the distance estimates we have reported would all be increased by 0.04
magnitudes.  We also note that the distances inferred for rapid rotators
are not consistent for the two colors; this implies that color temperatures
for these stars may be in error, especially if they are derived from \bv\
colors.  This may play some role in the lithium-rotation correlation seen in
young rapidly rotating stars.

We have shown that the internal 
consistency of MS fitting is high and, in the particular case of the open
clusters, the systematic errors are small.  The basic cluster data (abundances,
reddening, etc.) are also well established for the systems that we have 
studied in this paper.  The extremely good agreement between helioseismology
and theoretical solar models places strong constraints on missing physics in
the models, and by extension the properties of solar analogs should be 
accurately represented by the models.  For all of these reasons we believe
that the open cluster distance scale from MS fitting is on very strong ground.
    
The Hipparcos mission permits a comparison of parallax and MS fitting 
distances for a number of open clusters.  In two of the systems that we
have studied ($\alpha$ Per and the Hyades) the two distance scales
are in very good agreement.  In Coma and the Pleiades they disagree at the 
0.2 and 0.3 magnitude level respectively; these differences are at the 
3.4 and 3.7 $\sigma$ level respectively.  The different distance scales for
Praesepe are either very close (0.08 mag) or discrepant (0.33 mag) depending
upon which of the Hipparcos distance measurements is adopted; the latter
would be a 2 $\sigma$ disagreement.  We have 
searched for sources of error in the MS fitting distances of Coma Ber and
the Pleiades.  The $V-I$ photometry of Coma yields a distance that disagrees 
both with Hipparcos and \bv; this can be traced to a discrepancy in the 
temperature scales for the two colors in this cluster.  Although we
believe that there are a number of indications that the \bv\ temperature
scale is correct (consistency with spectroscopic temperatures and Stromgren
photometry, for example) reobserving this cluster in IR and near-IR colors 
would be highly desirable to quantify the magnitude of the problem.

In the case of the Pleiades there is no such ambiguity; different colors yield
identical distances.  Errors in the metal abundance and reddening as a solution
can be rejected on a variety of grounds.  The increase in the cluster helium 
abundance needed to reconcile the distance estimates is large and not 
consistent with direct measurements.  Furthermore, we can find no counterparts
of the Pleiades in the field, i.e. intrinsically faint solar abundance stars
(Soderblom et al. 1998).  We are therefore left with the uncomfortable 
choices of either requiring unknown physics in the interiors models or a
problem with the Hipparcos parallax distance scale to the Pleiades.  The
former choice is made even less attractive by the requirement that the models
retain agreement with the Sun, the other clusters, and numerous other tests
of the theory of stellar structure and evolution.  We therefore believe that
the latter explanation is more likely.

We have shown that there is evidence in the Pleiades data for
systematic errors in the parallaxes on small angular scales.  The same trends
are not present in the clusters where the two distance scales agree; they may
also be present in the Coma Ber cluster.  Clusters such as the Pleiades provide
many more stars within a small region of the sky than are present for the sky
as a whole, and they are therefore uniquely suited to test systematic effects
at small angular scales.  The other clusters and the Pleiades show no evidence
for systematic errors on scales larger than $1\deg$.  The Pleiades results
suggest that individual parallax measurements with large $\rho_\alpha^\pi$
correlations should be treated with caution.  The implications of this
result for other applications of the Hipparcos parallaxes will depend upon the
characteristics of the sample.  For large samples over large regions of the
sky the net effect will be a modest increase in the overall error.  A numerical
example:  Arenou et al. (1997, pp.441-443) find the overall ratio of
Hipparcos ``external'' to ``internal'' errors to be 1.06$\pm$0.07 from
clusters and 1.04$\pm$0.04 from distant stars.  With the internal error
$\sim$ 1 mas, this is equivalent to an additional (in quadrature) error
$\sim$ 0.2-0.4 mas.  This may in fact be the RMS size of the small-scale
errors.

\acknowledgments
D.R.S. acknowledges partial support from NASA grant NAGW-4837.  
R.B.H acknowledges partial support from NASA Grant NAG5-4830 and NSF grant
AST 9530632.  J.R.S. acknowledges partial support from NASA grants NAGW-2698
and NAG5-3363.  We wish to thank Debra Fischer, Eileen Friel, Burt Jones,
Jean-Claude Mermilliod, and Donald Terndrup for helpful discussions.

\newpage

\figcaption[fig1.eps]{The Pleiades compared with Praesepe (top panel) and 
$\alpha$ 
Per (bottom panel) in \bv.  Stars have been shifted by the Hipparcos 
distances in Table 1; photometry sources are listed in Section 2.  The 
Pleiades stars are the filled symbols and stars in the other clusters are the 
open symbols.}

\figcaption[fig2.eps]{The Pleiades compared with Praesepe (top panel) and 
$\alpha$ 
Per (bottom panel) in $(V-I)_C$.  Stars have been shifted by the Hipparcos 
distances in Table 1; Johnson and Kron $V-I$ have been converted to Cousins
$V-I$ as described in Section 2.  Photometry sources are listed in Section 
2.  The Pleiades stars are the filled symbols and stars in the other clusters 
are the open symbols.}

\figcaption[fig3.eps]{$M_V$ for a 1 Gyr as a function of log $T_{eff}$ (top
panel), \bv\ (middle panel) and $(V-I)_C$ (bottom panel).  In each panel the
bottom, middle, and top solid lines are [Fe/H]=-0.3,0, and +0.2 respectively;
the dashed line is a [Fe/H] = 0 isochrone with a helium abundance of 0.37 
(0.1 higher than the solar calibrated value).}

\figcaption[fig4.eps]{Isochrones with an age of 1 Gyr and a range of [Fe/H] 
are compared with field stars with $M_V>4$ and relative parallax errors less
than $5\%$ in \bv\ (top panel) and $(V-I)_C$ (bottom panel).  Age effects are
responsible for the departure of the data from the isochrone for $M_V<5$.
The top, middle, and bottom isochrones are respectively [Fe/H]=+0.2, 0, and
-0.3.}

\figcaption[fig5.eps]{Theoretical 600 Myr isochrones in the $M_{bol}/log T_{eff}$
plane for solar [Fe/H] (bottom lines) and Hyades [Fe/H] (top lines).  The 
solid lines are from this paper and the dashed lines are from Perryman et 
al. 1997.}

\figcaption[fig6.eps]{$M_V$ as a function of (\bv) (top panel) and $(V-I)_C$
(bottom panel) for Hyades members with Hipparcos parallax measurements and
(\bv)$>0.5$.  Photometry sources are listed in Section 2.  Solid points are 
single stars; open points are binaries from Griffin et al. 1988.  The 
empirical fits to the single star MS used in this paper are the solid lines.}

\figcaption[fig7.eps]{The difference between the Hipparcos $M_V$ and that
predicted from the isochrones for single Hyades stars with parallax 
measurements is plotted as a function of (\bv) (top panel) and $(V-I)_C$ 
(bottom panel).  The solid lines indicates the corrections as a function of 
color that were applied to the isochrones for consistency with the shape of
the Hyades MS.}

\figcaption[fig8.eps]{Histogram of differences between the Hipparcos $M_V$ and 
that predicted from the isochrones for (\bv) (top) and $(V-I)_C$ (bottom).
Data are binned in 0.1 magnitude intervals; single stars are the dark bins
and binaries are the light bins.}

\figcaption[fig9.eps]{Histogram of distance modulus estimates for Pleiades members,
binned in 0.05 magnitude intervals.  The isochrones were used for the top
panels, while the isochrones were corrected to the shape of the Hyades MS
in the bottom panels; those on the left side are distances from the (\bv)
color and those on the right side are distances from the $(V-I)$ color.
that predicted from the isochrones for (\bv) (top) and $(V-I)_C$ (bottom).
Stars with \bv\ from $0.5-0.75$ are the dark bins and stars with \bv\ from 
$0.76-0.9$ are the light bins.}

\figcaption[fig10.eps]{100 Myr theoretical isochrones with [Fe/H]=-0.03 
shifted 
to a distance of 5.60 are compared with the Pleiades in (\bv) (top) and 
$(V-I)$ (bottom).  Single stars are the filled squares, open circles 
are binaries from Bouvier, Rigaut, \& Nadeau 1997, and open circles with a
cross are rapid rotators.  The solid lines are the isochrones and the dashed 
lines are the same isochrones with the Hyades MS shape.}
  
\figcaption[fig11.eps]{As the top two panels of Figure 9, but with a wider 
color
range (\bv\ from $0.5-1.0$) and with data binned in 0.1 magnitude intervals.  
Single stars are the dark bins, binaries from Bouvier, Rigaut, \& Nadeau 
1997 are the light bins, and rapid rotators are the striped bins.}

\figcaption[fig12.eps]{As for Figure 9, except for Praesepe relative to a 
600 Myr isochrone.}

\figcaption[fig13.eps]{600 Myr theoretical isochrones with [Fe/H]=+0.04 
shifted to 
a distance of 6.16 are compared with Praesepe in (\bv) (top) and $(V-I)$ 
(bottom).  The solid lines are the isochrones and the dashed lines are the 
same isochrones with the Hyades MS shape.}

\figcaption[fig14.eps]{Histogram of distance modulus estimates for $\alpha$ 
Per 
members, binned in 0.05 magnitude intervals.  The top panel shows distances 
from the (\bv) color and those on the bottom panel are distances from the 
$(V-I)$ color; isochrones with the Hyades MS shape yield similar results and
are not shown.  Slow rotators  are the dark bins and rapid rotators 
(v sin i $> 49$ km/s) are the light bins.}

\figcaption[fig15.eps]{50 Myr theoretical isochrones with [Fe/H]=-0.05 (solid 
lines) shifted to a distance of 6.23 are compared with $\alpha$ Per in 
(\bv) (top) and $(V-I)$ (bottom).  Slow rotators are the filled squares and
rapid rotators are the open circles with a cross.}

\figcaption[fig16.eps]{Histogram of distance modulus estimates for Coma Ber 
members, binned in 0.05 magnitude intervals.  The top panel shows distances 
from the (\bv) color and those on the bottom panel are distances from the 
$(V-I)$ color.  Stars with (\bv) from $0.5-0.9$ are the dark bins, which 
corresponds to the color interval used for the other clusters.  The light
bins are hotter stars with (\bv) from $0.35-0.499$.  Note that for the
cooler stars there is no well-defined peak for $(V-I)$, and that the overall
peak disagrees significantly with the Hipparcos and (\bv) distances (see
text).}

\figcaption[fig17.eps]{Effective temperatures for Coma Ber members from 
Boesgaard (1987) compared with those predicted from (\bv) (top) and 
$(V-I)$ (bottom).}

\figcaption[fig18.eps]{500 Myr theoretical isochrones (solid lines) with 
[Fe/H]=-0.07 shifted to a distance of 4.54 are compared with Coma Ber 
in (\bv) (top) and $(V-I)$ (bottom).}

\figcaption[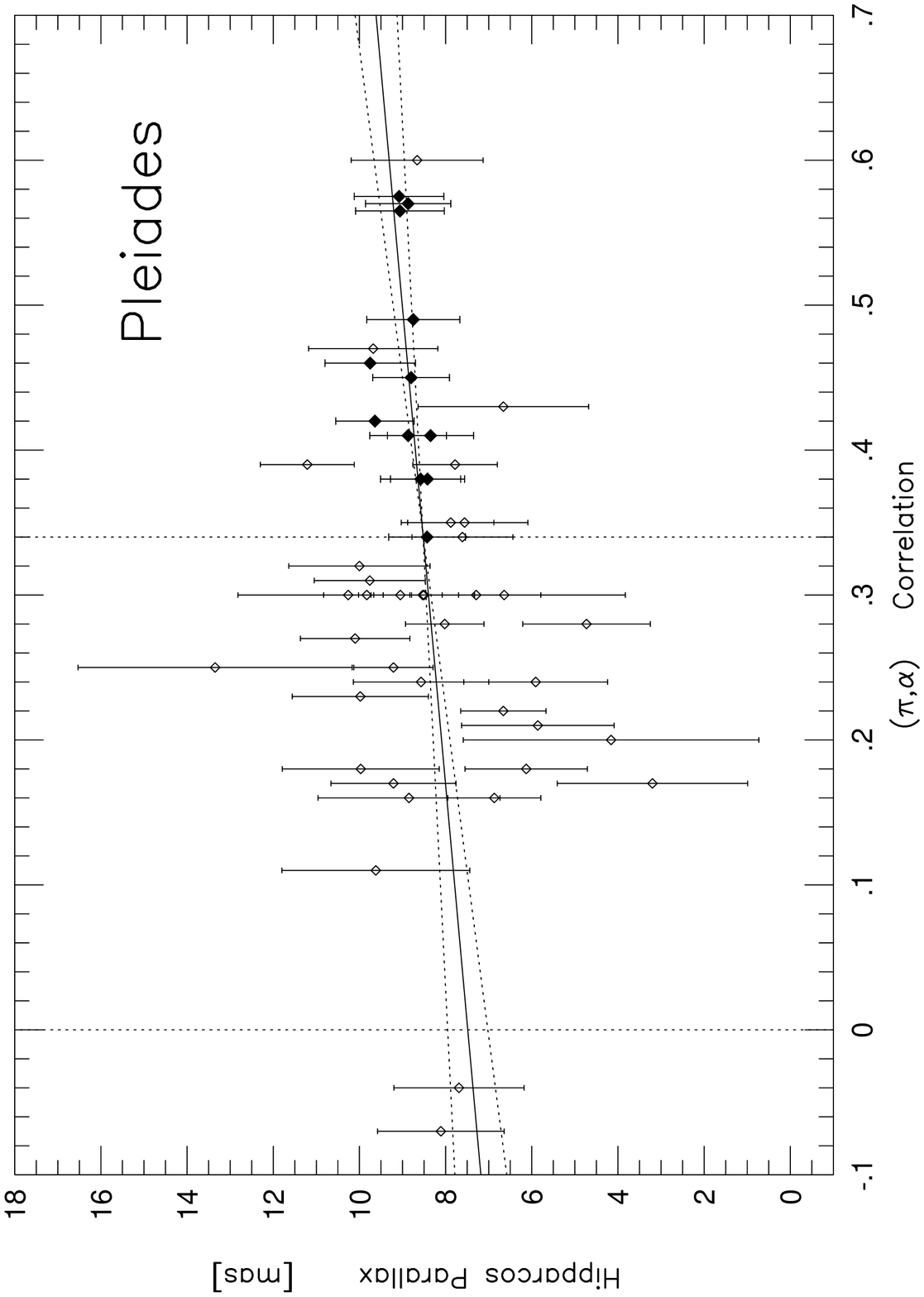]{Hipparcos parallax vs. the correlation 
$\rho_\alpha^\pi\ $ for 49 Pleiades 
members.  Filled symbols are 12 bright ($V < 7$) stars within $\sim 1 \deg$
of the cluster center with correlations $\rho_\alpha^\pi \geq +0.34$.
Vertical dotted lines mark $\rho_\alpha^\pi = 0$ and the mean value $+0.34$.
Sloping lines represent the weighted least-squares relation 
$\pi \ = \ 8.53 \ + \ (3.04 \pm 1.36)(\rho_\alpha^\pi - 0.34)$ mas.}

\figcaption[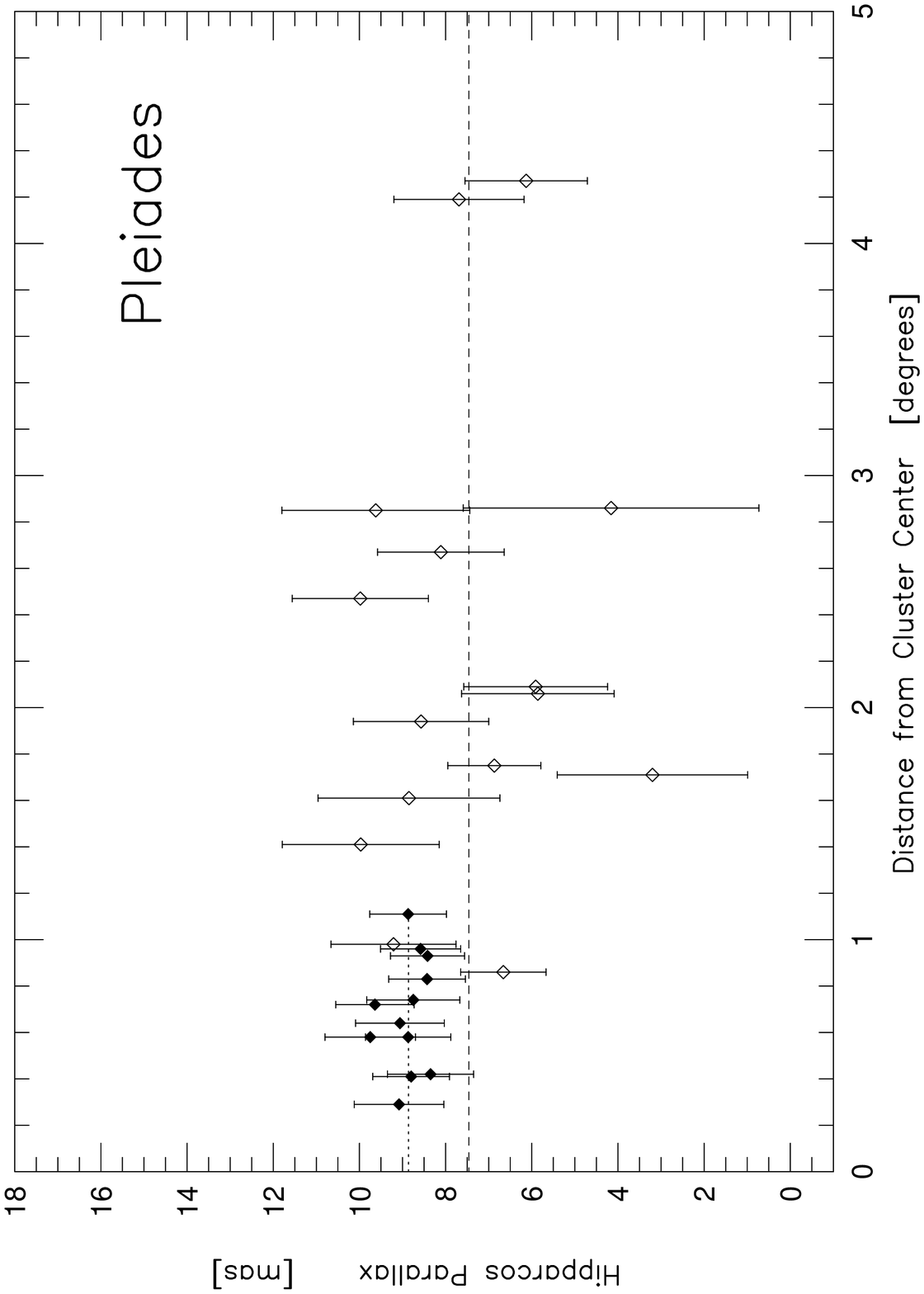]{Hipparcos parallax vs. angular distance from the 
Pleiades cluster center.  
The filled symbols are the same 12 bright stars with high $\rho_\alpha^\pi$ 
as in Fig. 14.  The open symbols are the 15 stars with $\rho_\alpha^\pi 
< +0.25$, with no restriction on magnitude or distance.  The long-dashed
line marks the mean parallax (7.46 mas) for these 15 stars.
The dotted line marks the mean parallax (8.86 mas) for the 12 bright stars.}

\figcaption[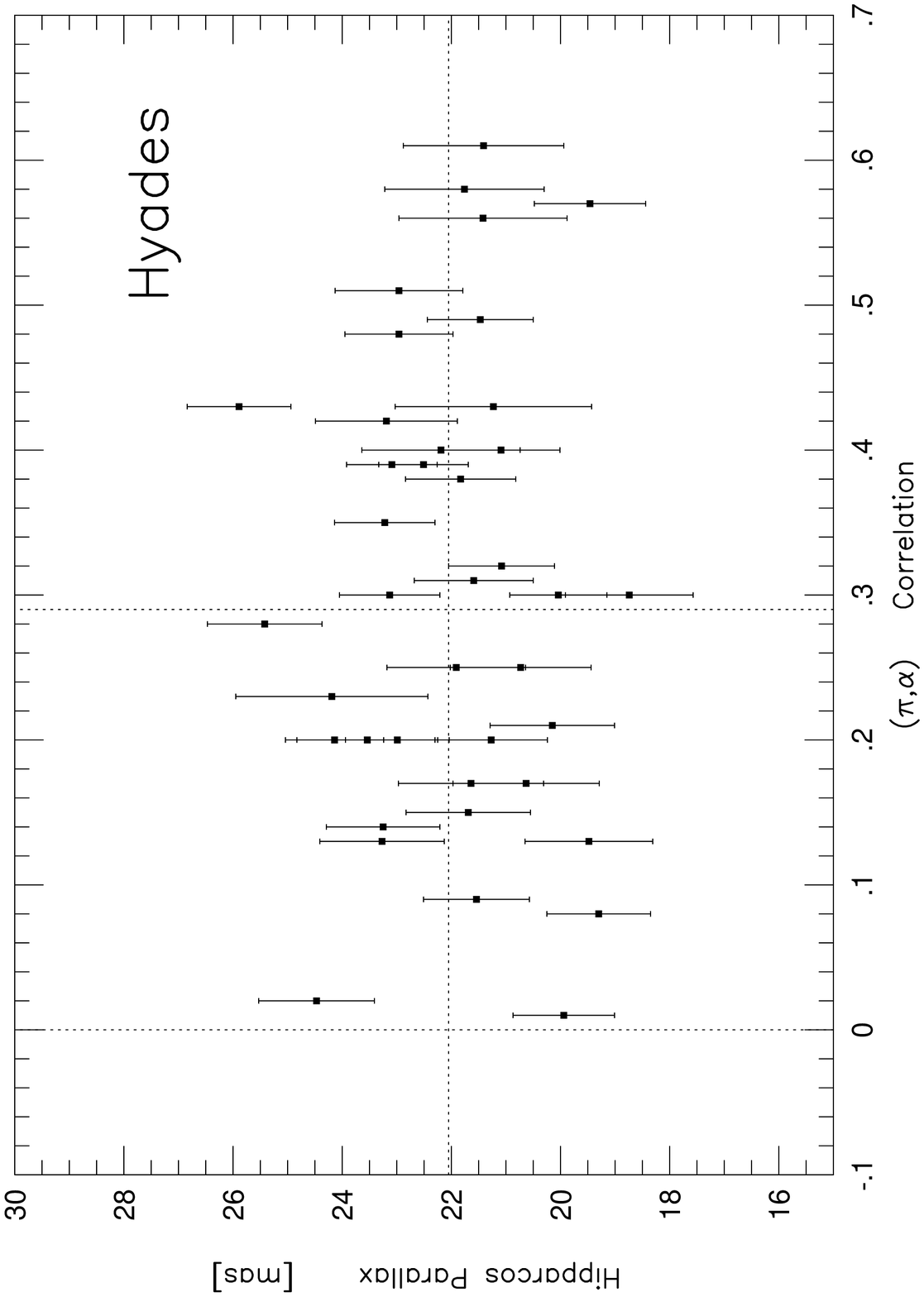]{Hipparcos parallax vs. the correlation 
$\rho_\alpha^\pi\ $ for the 40 Hyades 
cluster members in Table 8 of Perryman et al. (1997).
Vertical dotted lines mark $\rho_\alpha^\pi = 0$ and the mean value $+0.29$.
Horizontal dotted line marks the mean parallax 22.05 mas.}

\figcaption[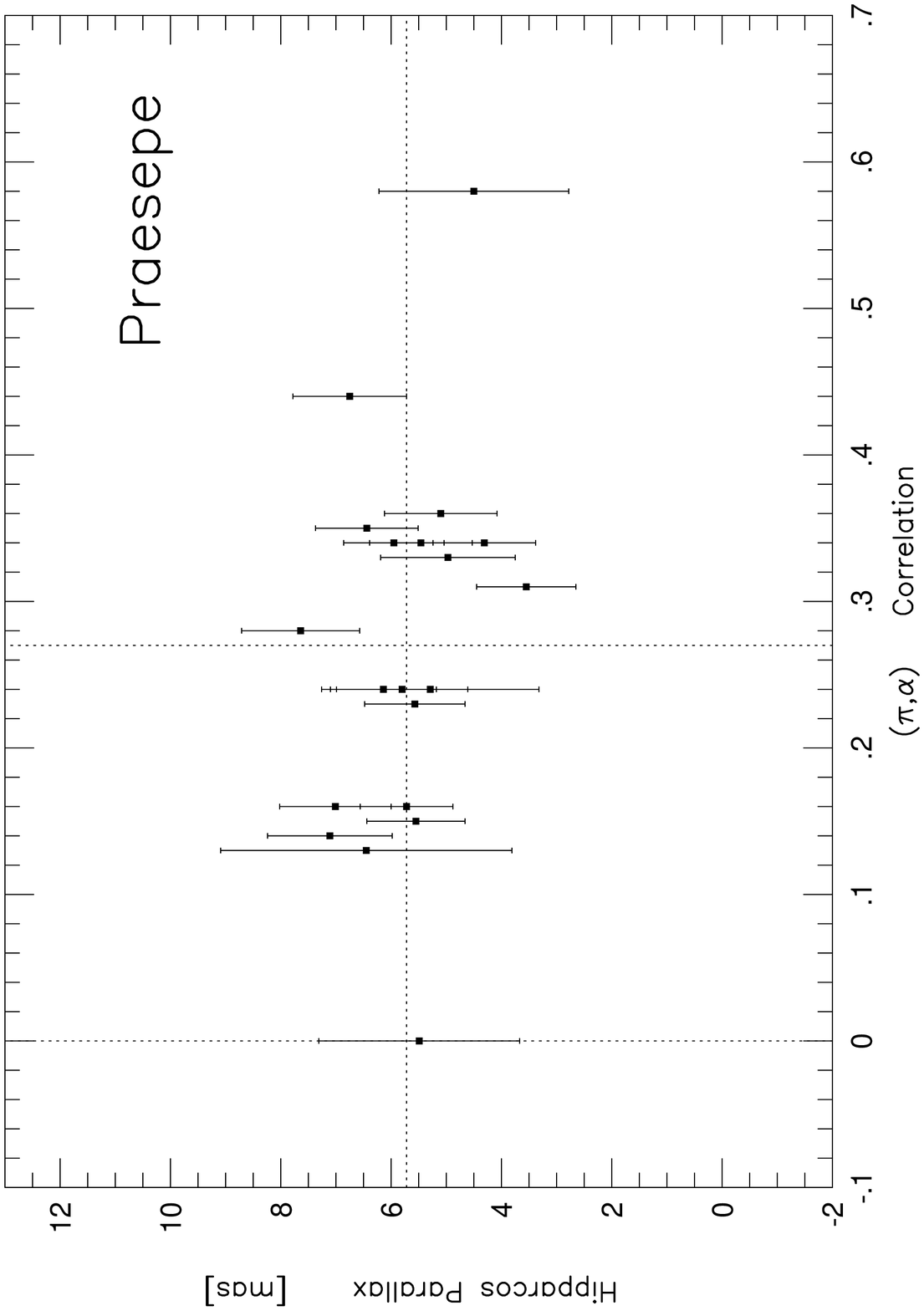]{Hipparcos parallax vs. the correlation 
$\rho_\alpha^\pi\ $ for 20 Praesepe 
cluster members verified by proper motion and position in the color-magnitude 
diagram.  Vertical dotted lines mark $\rho_\alpha^\pi = 0$ and the mean value 
$+0.27$.  Horizontal dotted line marks the mean parallax 5.72 mas.}
 
\figcaption[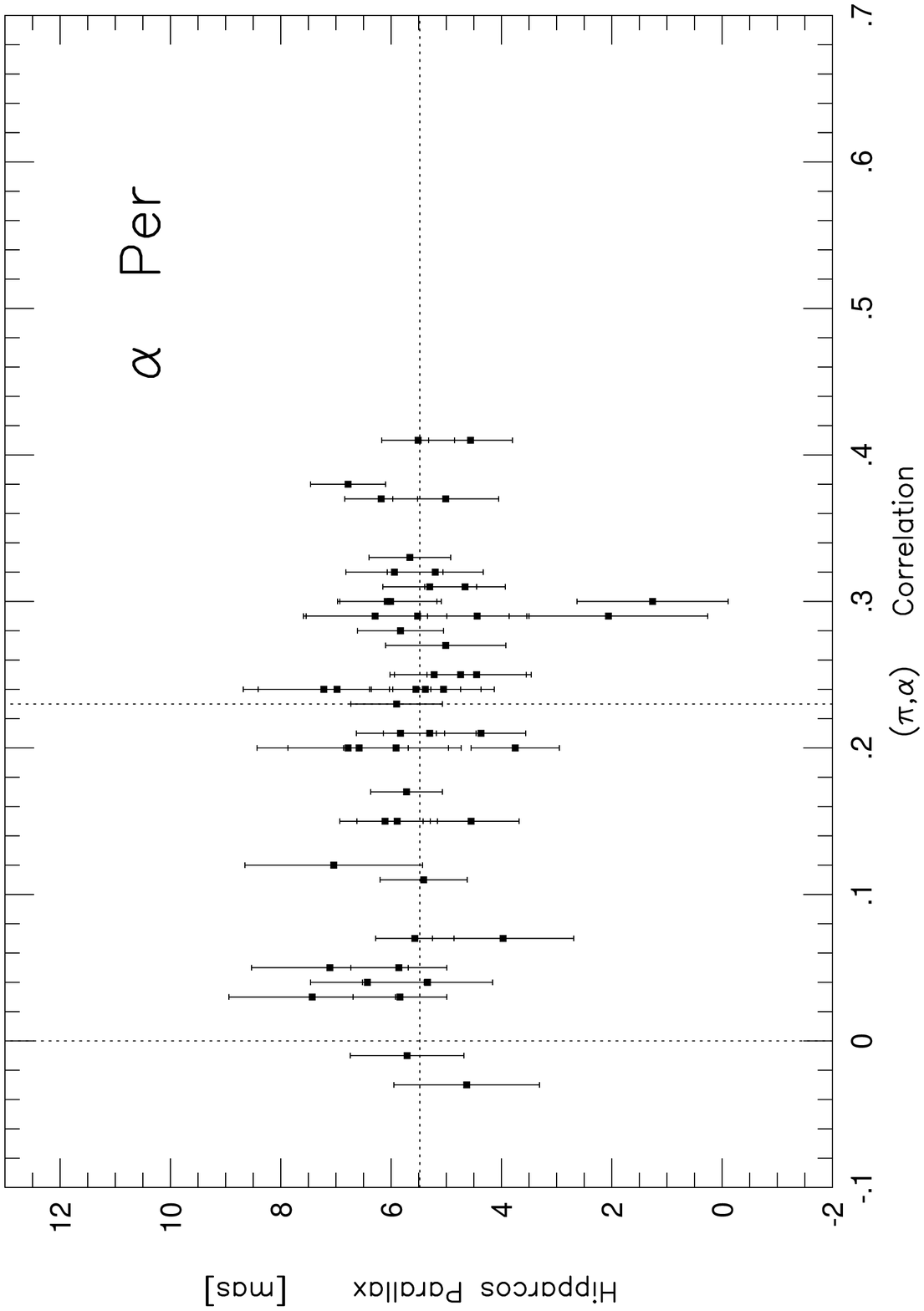]{Hipparcos parallax vs. the correlation 
$\rho_\alpha^\pi\ $ for 51 $\alpha$~Per
cluster members verified by proper motion, radial velocity, and position in 
the color-magnitude diagram.  Vertical dotted lines mark $\rho_\alpha^\pi = 0$
and the mean value $+0.23$.  Horizontal dotted line marks the mean parallax 
5.48 mas.}

\figcaption[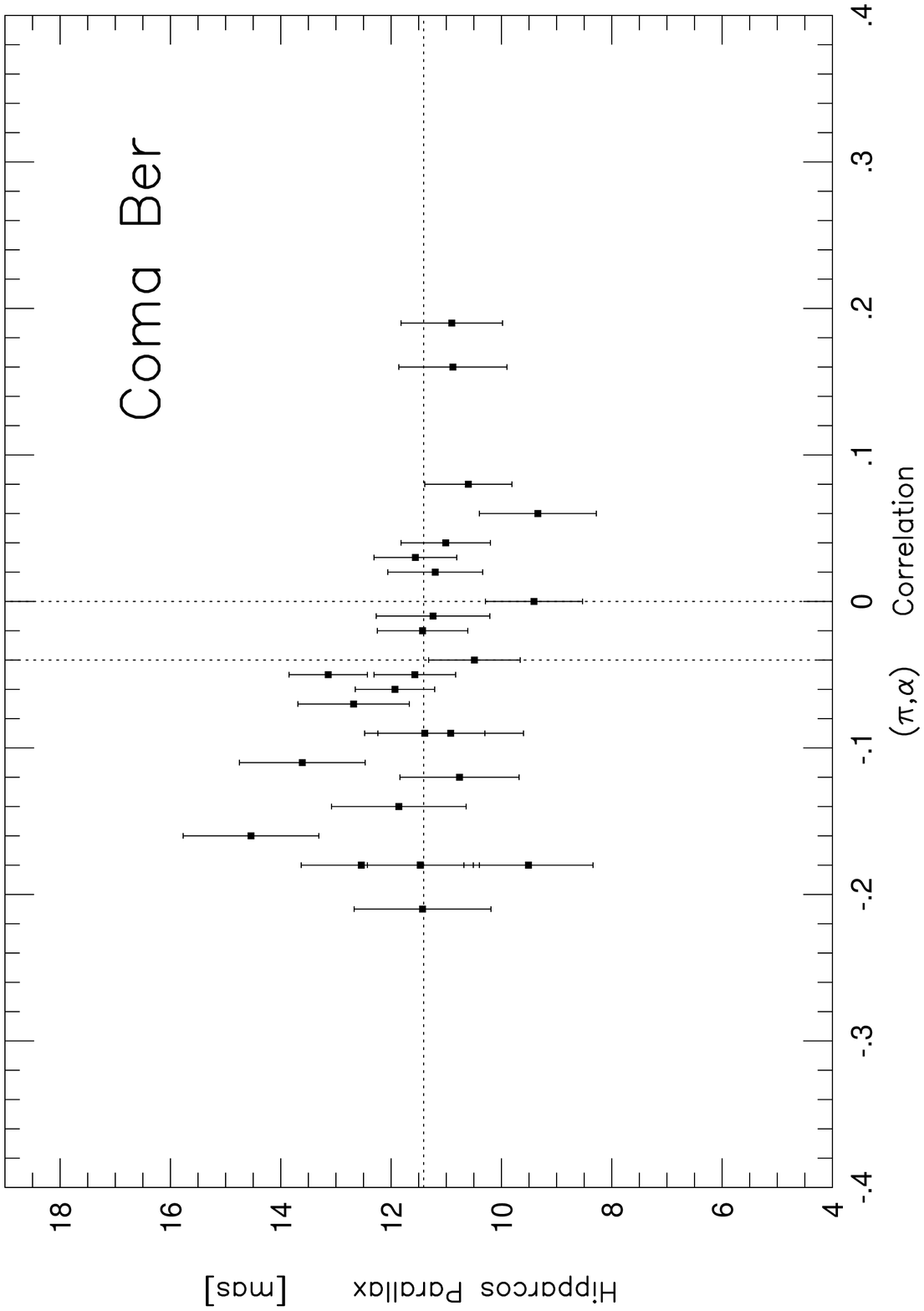]{Hipparcos parallax vs. the correlation 
$\rho_\alpha^\pi$ for 26 Coma~Ber
cluster members verified by proper motion and position in the color-magnitude 
diagram.  Vertical dotted lines mark $\rho_\alpha^\pi = 0$ and the mean value 
$-0.04$.  Horizontal dotted line marks the mean parallax 11.41 mas.}

\end{document}